\documentclass[
    floatfix,
    aps,
    prd,
    a4paper,
    twocolumn,
    shortbibliography,
    notitlepage,
    superscriptaddress
]{revtex4-1}

\usepackage[utf8]{inputenc}

\usepackage{amsmath,amsthm,amssymb}
\usepackage{bm}
\usepackage{siunitx}

\usepackage{centernot}

\usepackage{graphicx}
\usepackage{xcolor}

\usepackage[colorlinks=true]{hyperref}

\usepackage[caption=false,labelformat=simple,listofformat=subsimple]{subfig}

\usepackage{etoolbox}

\usepackage{csquotes}
\usepackage{microtype}

\usepackage[capitalise]{cleveref}
\crefname{equation}{}{}

\newcommand{\LCp}{{\scriptscriptstyle +}}
\newcommand{\LCpm}{{\scriptscriptstyle \pm}}

\newcommand{\LCperp}{{\scriptscriptstyle \perp}}

\newcommand{\ud}{\ensuremath{\mathrm{d}}}
\newcommand{\cE}{\ensuremath{\mathcal{E}}}
\newcommand{\CA}{\ensuremath{\mathcal{A}}}
\newcommand{\CB}{\ensuremath{\mathcal{B}}}

\DeclareMathOperator{\Ai}{Ai}
\DeclareMathOperator{\Bi}{Bi}
\DeclareMathOperator{\Ei}{Ei}

\newcommand{\ordo}[1]{\ensuremath{%
        \mathcal{O}\big( #1 \big )
    }
}

\newcommand{\LL}[1]{\ensuremath{\text{LL}_{#1}}}

\newcommand{\cemph}[1]{#1}

\begin{document}

\title{ Reduction of Order and Transseries Structure of Radiation Reaction }
\author{Robin Ekman}
\email{robin.ekman@umu.se}
\affiliation{ Centre for Mathematical Sciences, University of Plymouth, Plymouth, PL4 8AA, UK }
\affiliation{ Department of Physics, Umeå University, SE-901 87 Umeå, Sweden }
\date{\today}

\begin{abstract}
    The Landau-Lifshitz equation is obtained from the Lorentz-Abraham-Dirac equation through `reduction of order'.
    It is the first in a divergent series of approximations that, after resummation, eliminate runaway solutions.
    Using Borel plane and transseries analysis we explain why this is, and show that a non-perturbative formulation of reduction of order can retain runaway solutions.
    We also apply transseries analysis to solutions of the Lorentz-Abraham-Dirac equation, essentially treating them as expansions in both time and a coupling.
    Our results illustrate some aspects of such expansions under changes of variables and limits.
\end{abstract}
\maketitle

\section{Introduction}
\label{sec:intro}

Radiation reaction (RR) continues to attract attention in classical and quantum electrodynamics, both experimentally~\cite{Cole:2017zca,Poder:2018ifi} and theoretically~\cite{Burton:2014wsa,Blackburn:2019rfv,Gonoskov:2021hwf} with a particular focus on intense laser fields where RR forces compare to or dominate the Lorentz force ~\cite{danson2019petawatt,Abramowicz:2021zja,Meuren:2020nbw}.
RR in strong fields is also relevant in gravitational physics, first clearly observed in the Hulse-Taylor binary pulsar~\cite{Taylor:1979zz}, and studied theoretically in, e.g., Refs.~\cite{Damour:2020tta,DiVecchia:2021bdo,Herrmann:2021tct,Bjerrum-Bohr:2021vuf}.

Recently many authors have applied resummation~\cite{Torgrimsson:2020mto,Torgrimsson:2020wlz,Karbstein:2019wmj,Mironov:2020gbi,Heinzl:2021mji,Torgrimsson:2021wcj,Ekman:2021eqc,Torgrimsson:2021zob} and
resurgence and transseries concepts~\cite{Taya:2020dco,Taya:2021dcz,Dunne:2021acr} in classical and quantum electrodynamics in strong backgrounds.
(For introductions to and reviews of these concepts, see Refs.~\cite{Bender:1999,Marino:2012zq,Dorigoni:2014hea,Dunne:2015eaa,Aniceto:2018bis,Costin:2019xql,Costin:2020hwg}.)
As a prominent example all-orders, resummed results~\cite{Mironov:2020gbi,Heinzl:2021mji} have been vital to progress on the Ritus-Narozhny conjecture~\cite{PhysRevD.21.1176,Fedotov:2016afw} of the breakdown of Furry picture perturbation theory

In this paper we use the the Lorentz-Abraham-Dirac (LAD) equation of motion for radiation reaction~\cite{abraham1905,lorentz1909,dirac1938classical} as a ``test-bed'' for transseries analysis.
It is a natural choice of a simple setting in which to explore transseries structures, essentially because we know that they must be there and their physical interpretation.
They are the `unwanted' features of the LAD equation, pre-acceleration and runaway solutions, that are explicitly non-perturbative in $\tau_0$, the time-scale of radiation reaction.
Indeed these are not seen in perturbative approaches, including reductive procedures which lead to e.g.~the Landau-Lifshitz~\cite{LandauLifshitzII} (LL) equation, at any order~\cite{Ekman:2021eqc}.
We will see that the time-dependent nature of our problem means that even though the physics is quite simple, the formal structure can still be rich.

We extend our previous work~\cite{Ekman:2021eqc}, which iterated `reduction of order' \emph{ad infinitum} in a constant crossed field (CCF) to obtain the all-orders (in $\tau_0$) equation of motion $\LL\infty$ by showing that this procedure eliminates non-perturbative transseries structure at the level of the equation of motion.
We also show that the same holds in a circularly polarised monochromatic plane wave.
The elimination of non-perturbative terms is, however, dependant on an ``initial condition'' matching to the Lorentz force at vanishing field.
Other ``initial conditions'' keep non-perturbative terms and lead to runaway solutions of the order-reduced equation of motion.
We then consider inserting a hard cutoff into a constant field;
this is the simplest time-dependence which allows us to unambiguously investigate pre-acceleration and its transseries structure.

This paper is organised as follows.
We begin in \cref{sec:LLINF} by reviewing reduction of order as applied to the LAD equation, and $\LL\infty$.
We show that non-pertubative contributions to $\LL\infty$ are large as the coupling goes to zero, and lead to runaway solutions if kept.
Next, in \cref{sec:step} we solve the two equations of motion in a step field profile, finding on the level of solutions to LAD
instanton terms that are precisely the pre-accelerating and runaway solutions.
We conclude in \cref{sec:concs}.

\section{\texorpdfstring{$\LL\infty$}{LL∞}: reduction of order and transseries}
\label{sec:LLINF}

\subsection{Conventions and notations}

We will consider the momentum $p^\mu$ of a particle of charge $e$ and mass $m$ in a constant crossed field (CCF) given by
\begin{equation}
    f_{\mu\nu} := \frac{e}{m} F_{\mu\nu} = \cE m \, n_{[\mu} \epsilon_{\nu]}
\end{equation}
where $\cE$ is the dimensionless field strength, $n_\mu$ is lightlike and $\epsilon^2 = -1$ with $n \cdot \epsilon = 0$.
As we will only be concerned with one species of particle we henceforth use units where $m = 1$, although we will restore $m$ in places for clarity.
We will use lightfront coordinates $p^\LCpm = p^0 \pm p^z, p^\LCperp = (p^\mathfrak{1}, p^\mathfrak{2})$, the $z$-axis aligned such that $p^\LCp = n \cdot p$.

The Lorentz-Abraham-Dirac (LAD) equation reads, using an overdot for derivative with respect to proper time,
\begin{equation}
    \label{eq:LAD}
    \dot{p}_\mu = f_{\mu\nu}p^\nu + \tau_0 P_{\mu\nu} \ddot{p}_\rho
\end{equation}
where $P_{\mu\nu} = g_{\mu\nu} - p_\mu p_\nu$ is the projector orthogonal to $p_\mu$ and
\begin{equation}
    \tau_0 := \frac{2\alpha}{3m}
    \, ,
\end{equation}
$\alpha$ being the fine-structure constant;
for an electron $\tau_0 \approx \SI{6.2e-24}{s}$.
\cemph{Non-perturbative effects in the solutions to LAD occur on time-scales of $\tau_0$,
    but radiation reaction has observable effects over much longer time-scales.
}
The interaction is characterised by an energy parameter,
\begin{equation}
    \delta^2  = \tau_0^2 p_\mu f^{\mu\nu} f_{\nu\rho} p^\rho = ( \tau_0 \cE p^\LCp)^2
    \,
    .
\end{equation}
When working at the level of the solution, the initial value $\delta_0$ will play the role of a coupling.
Note that $\delta = \tau_0 \chi$ where $\chi$ is the quantum non-linearity parameter~\cite{DiPiazza:2011tq,Gonoskov:2021hwf};
they are related in the same way that the classical electron radius and the Compton length are.
\cemph{
    This means that values $\delta \gtrsim 1$ are deeply in the quantum regime, and mainly relevant for classical electrodynamics as a formal theory.
}

\subsection{Reduction of order and \texorpdfstring{$\LL\infty$}{LL∞}}

The Landau-Lifshitz (LL) equation~\cite{LandauLifshitzII} is obtained from Lorentz-Abraham-Dirac by reduction of order:
we apply $\ud/\ud \tau$ to both sides of~\cref{eq:LAD}, substitute for $\dot{u}$ according to~\cref{eq:LAD} itself, and discard terms of order $\tau_0^2$.
This yields, in general,
\begin{equation}
    \label{eq:LL1}
    \dot{p}_\mu = f_{\mu\nu} p^\nu + \tau_0 \left[
        (P f^2)_{\mu\nu} p^\nu + p^\rho \partial_\rho f_{\mu\nu} p^\nu
    \right]
    \,
    ,
\end{equation}
although the final, gradient, term of course vanishes for a CCF.

The reduction of order procedure as just described reduces the order \emph{in time}, but the procedure can be iterated any number of times to any order \emph{in $\tau_0$}~\cite{PiazzaExact,Ekman:2021vwg}.
We will therefore refer to the first iteration~\cref{eq:LL1} as $\LL1$.
If reduction of order is iterated \emph{ad infinitum}, i.e., to all orders in $\tau_0$, it yields the equation of motion $\LL\infty$,
\begin{equation}
    \label{eq:LL-infinity}
    \dot{p}^\mu = \mathcal{A}(\delta)f^{\mu\nu}p_\nu + \tau_0 \mathcal{B}(\delta)(Pf^2)^{\mu\nu}p_\nu
    \,
    ,
\end{equation}
as discussed in a previous paper~\cite{Ekman:2021eqc}.
Here the functions $\mathcal{A}$ and $\mathcal{B}$ are solutions of the ODE:s
\begin{equation}
    \label{eq:fixed-point}
    \left\{
        \begin{array}{rcl}
            \delta^3 \CB \frac{\ud \CA}{\ud \delta} &=& 1 - \CA - 2 \delta^2 \CA \CB \\ % [10pt]
            \delta^3 \CB \frac{\ud \CB}{\ud \delta} &=&   - \CB - 2 \delta^2 \CB^2 + \CA^2 \\ % [10pt]
        \end{array}
    \right.
    \,
    .
\end{equation}
and the initial conditions that recover first-order Landau-Lifshitz are
\begin{equation}
    \label{eq:A-B-ic}
    \CA(0) = \CB(0) = 1 \, .
\end{equation}
The functions $\CA, \CB$ encode how the RR force varies with energy, vaguely analogous to a running coupling.

We emphasise here that when $\CA, \CB$ verify~\cref{eq:fixed-point} the solution of $\LL\infty$ is a solution of LAD.
Explicitly, differentiating~\cref{eq:LL-infinity} we obtain
\begin{widetext}
    \begin{equation}
        \ddot{p}_\mu
        = \frac{\ud \CA}{\ud \delta} \frac{\ud \delta}{\ud \tau} f_{\mu\nu} p^\nu + \CA f_{\mu\nu} \dot{p}^\nu
         + \tau_0 \left[
             \frac{\ud \CB}{\ud \delta} \frac{\ud \delta}{\ud \tau} (P f^2)_{\mu\nu} p^\nu
             + \CB \left(
                 (P f^2)_{\mu\nu} \dot{p}^\nu
                 - \dot{p}_\mu (p^\LCp \cE)^2
                 - p_\mu (\dot{p} f^2 p)
             \right)
         \right]
         \,
         .
    \end{equation}
    Now dotting $n^\mu$ into LAD, it reads
    \begin{equation}
        \label{eq:LL-inf-derivation}
        n \cdot \dot{p} = \tau_0 ( n \cdot \ddot{p} - p^\LCp p\cdot \ddot{p} )
        =
        - \tau_0^2 \left[
            \frac{\ud \CB}{\ud \delta} \frac{\ud \delta}{\ud \tau} (p^\LCp)^3 \cE^2
            + 2 \CB p^\LCp (\dot{p} f^2 p) + \CB (n \cdot \dot{p}) (p^\LCp \cE)^2
        \right]
        - \tau_0 \CA p^\LCp (p f \dot p) - \tau_0^2 p^\LCp \CB (\dot{p} f^2 p)
        \,
        .
    \end{equation}
    It follows from~\cref{eq:LL-infinity} that $p f \dot{p} = \CA (p^\LCp \cE)^2$ and $ \dot{p} f^2 p = -\tau_0 \CB (p^\LCp \cE)^4$;
    we also have $ \frac{\ud \delta}{\ud \tau} = \tau_0 \dot{p}^\LCp \cE = -\tau_0^2 \CB (p^\LCp \cE)^3$.
    Substituting these into the RHS of~\cref{eq:LL-inf-derivation}, writing out the LHS according to~\cref{eq:LL-infinity}, and dividing by $\tau_0 (p^\LCp)^3 \cE^2$ it becomes
    \begin{equation}
        -\CB
        =
        (\tau_0 p^\LCp \cE)^3 \CB \frac{\ud \CB}{\ud \delta}
        + 2 (\tau_0 p^\LCp \cE)^2 \CB^2
        - \CA^2
        \,
        ,
    \end{equation}
    which is one of the ODE:s~\cref{eq:fixed-point}.
    Hence the ${}^\LCp$ component of LAD will be satisfied if~\cref{eq:LL-infinity} holds, with $\CB$ a solution to~\cref{eq:fixed-point}.
    A similiar calculation shows that the transverse components of LAD will be satisfied if~\cref{eq:LL-infinity} holds and $\CA$ is a solution to~\cref{eq:fixed-point}.
    The remaining component is fixed by the mass-shell condition.
\end{widetext}

As $\LL\infty$ is obtained from reduction of order in a small parameter, it is essentially a resummed perturbative expansion. It is therefore entirely possible that the procedure could miss non-perturbatively small terms in the expansion parameter. We will here investigate the possible presence of such terms.

That is, \cref{eq:fixed-point,eq:A-B-ic} can be solved as perturbative series $\CA \sim 1 - 2 \delta^2 + \ldots, \CB \sim 1 - 6 \delta^2 + \ldots$.
Although divergent, these series can be resummed with the Borel-Pad\'e (see, e.g., Ref.~\cite[Ch.~8]{Bender:1999}) or ``educated match''~\cite{Alvarez:2017sza} methods.
\cemph{
    As pointed out in Ref.~\cite{Ekman:2021eqc}, $\LL\infty$ remains causal and free of runaways after such a resummation of perturbative terms.
    This is in contrast to the non-relativistic case studied in Ref.~\cite{Zhang:2013ria}, where these non-perturbative effects appear precisely after performing a Borel resummation.
    The question is thus raised whether non-perturbative effects appear from solutions of a more general transseries form
}
\begin{equation}
    \label{eq:trans-ansatz}
    \begin{Bmatrix}
        \CA \\ \CB
    \end{Bmatrix}
    \sim \sum_{k, \ell \ge 0}
    \begin{Bmatrix}
        A_{k, \ell} \\
        B_{k, \ell}
    \end{Bmatrix}
    \delta^{2k} e^{-\ell \kappa / \delta^\lambda}
    \,
    ,
\end{equation}
(for some $\kappa, \lambda$ to be determined)
which are not found by perturbative expansion or numerics.
We use $\sim$ rather than equality here and treat, for now, the expansion~\cref{eq:trans-ansatz} formally -- the space of such transseries is closed under algebraic operations and differentiation.

To determine the parameters $\kappa, \lambda$ we linearise around $(\CA, \CB) = (1,1)$ and $\delta = 0$;
the general solution of the linearisation is
\begin{subequations}
    \label{eq:gen-sol}
    \begin{align}
        \CA & = 1 - 2 \delta^2 + c_1 \frac{1}{\delta^2} e^{1/2\delta^2} + \ordo{\delta^3} \\
        \CB & = 1 - 6 \delta^2 + c_1 \frac{1}{\delta^4} e^{1/2\delta^2} + c_2 \frac{1}{\delta^2} e^{1/2\delta^2} + \ordo{\delta^3}
    \end{align}
\end{subequations}

for arbitrary constants $c_1, c_2$.
We see that there are indeed non-perturbative terms depending exponentially on $1/\delta^2$, but these are \emph{large} for real $\delta$.
The only solution finite as $\delta \searrow 0$ has $c_1 = c_2 = 0$, and hence lacks a non-perturbative part (its perturbative expansion is, we stress, divergent and must be resummed, though).
We return to this at the end of this Section.

We can strengthen our argument through the interpretation of $\kappa = -1/2$ as the location of the convergence-limiting singularity in the complex Borel plane.
Borel singularities, and the overall transseries structure, are intimately related to the large-order growth of the perturbative coefficients~\cite{Borinsky:2021hnd}.
In our case this can be determined to be, to leading order,
\begin{equation}
    A_{k}, B_{k} \sim (-2)^k k!
\end{equation}
by computing many coefficients using the recursion relations in Ref.~\cite{Ekman:2021eqc}.
We compute a normalised Borel transform
\begin{equation}
    \label{eq:Borel}
    \operatorname{Borel}[\CA](t)
    =
    \sum_k \frac{A_{k}}{2^k k!} t^k
    \,
    .
\end{equation}
The transform cancels the factorial growth of the $A_n$, producing a series with finite radius of convergence, which can be analytically continued.
With this normalisation we expect the leading singularity to appear at $t = -1$.

\begin{figure}[btp]
    \centering
    \includegraphics[width=\linewidth]{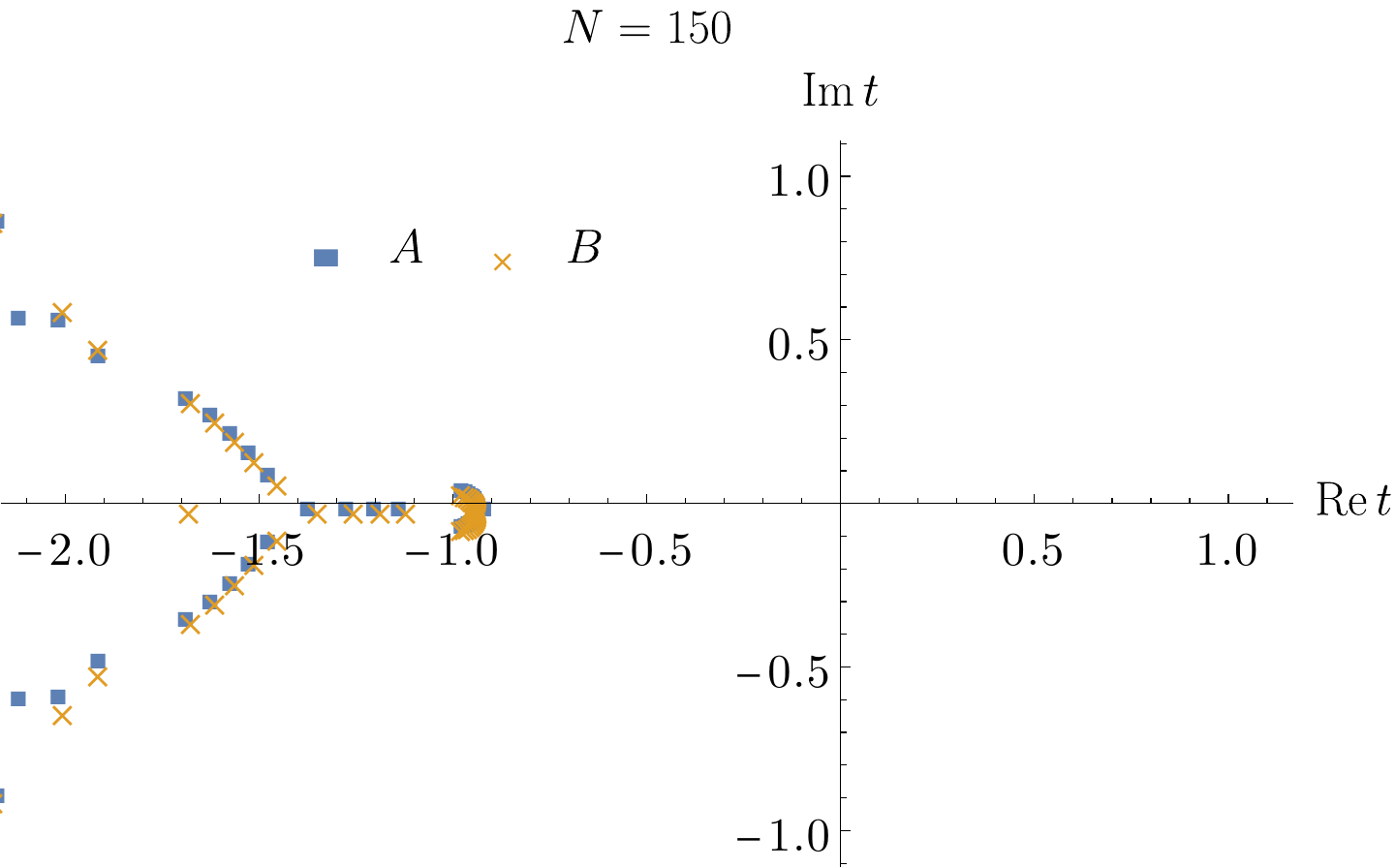}
    \caption{
        Borel-Pad\'e poles accumulating along the negative real axis, indicating the presence of a branch cut.
    }
    \label{fig:poles}
\end{figure}

The convergence-limiting singularity of the analytical continuation can now be probed using Pad\'e approximants.
The Pad\'e method can struggle to identify multiple branch cuts, as it must accumulate poles along a cut to approximate it.
This difficulty can be circumvented with a conformal map~\cite{Kleinert:2001ax,le2012large,Costin:2019xql,Costin:2020hwg}, making it also possible to identify singularities beyond the leading~\cite{Borinsky:2021hnd,Dunne:2021acr} and increase the accuracy of resummations~\cite{Florio:2019hzn,Torgrimsson:2020wlz,Costin:2021bay}
Even without conformal mapping, though, there is a clear accumulation of Borel-Pad\'e poles along the ray $t \le -1$, seen in \cref{fig:poles}.

\begin{figure}[tbp]
    \centering
    \includegraphics[width=\linewidth]{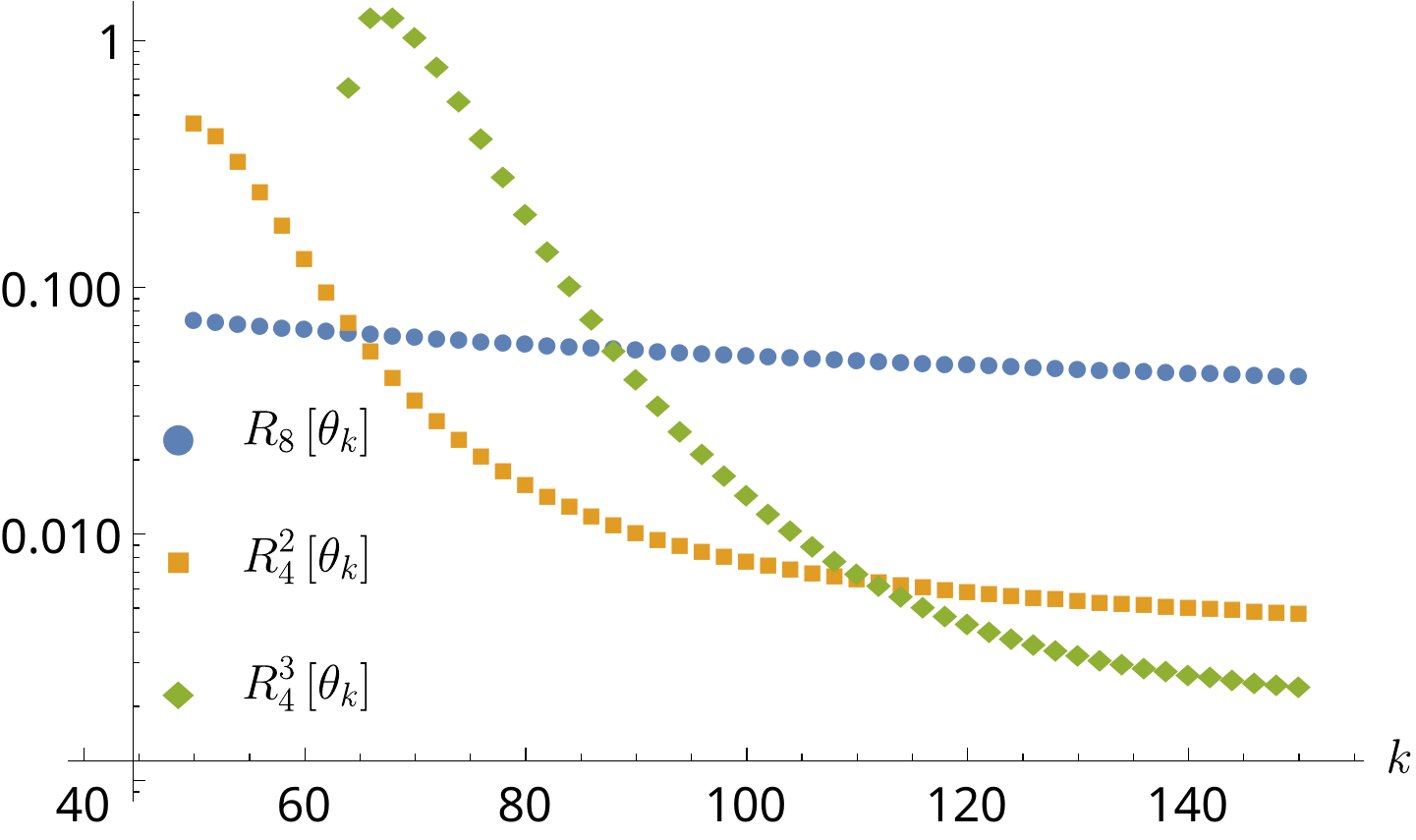}
    \caption{
        Slow convergence of $\theta_k = -\frac{B_k}{k B_{k-1}} - 2$ as $k \to \infty$, even applying order $8$ Richardson extrapolation ($R_8$), due to subleading logarithmic corrections.
        The modified extrapolations $R^{(2,3)}_K$ are accurate up to order $(\log k)^{(1,2)}/n^{-K}$.
    }
    \label{fig:richardson}
\end{figure}

A fairly large number of terms are needed to see the structure in \cref{fig:poles}.
The reason for this is that while
\begin{equation}
    \frac{B_k}{k B_{k-1}} \xrightarrow{k \to \infty} \frac{1}{\kappa} = -2
    \,
    ,
\end{equation}
there are slowly decaying subleading corrections.
Even after applying high-order Richardson extrapolation~\cite[Ch.~8.1]{Bender:1999}, the slow convergence persists.
Experimentally, this is because the subleading large-order behaviour of the coefficients is \emph{logarithmic}
\begin{equation}
    \frac{B_k}{k B_{k-1}} \approx -2 \left[
        1 + \frac{\Lambda}{k} (\log k)^2 + \mathcal{O}\left( (\log k)^2/k^2 \right)
    \right]
    \,
    ,
\end{equation}
and so not eliminated by standard Richardson extrapolation.
Modifying Richardson extrapolation to account for logarithmic corrections (see Ref.~\cite{Borinsky:2021hnd} and \cref{app:richardson}), the convergence is improved significantly, as shown in \cref{fig:richardson}.

\begin{figure}[tbp]
    \centering
    \includegraphics[width=\linewidth]{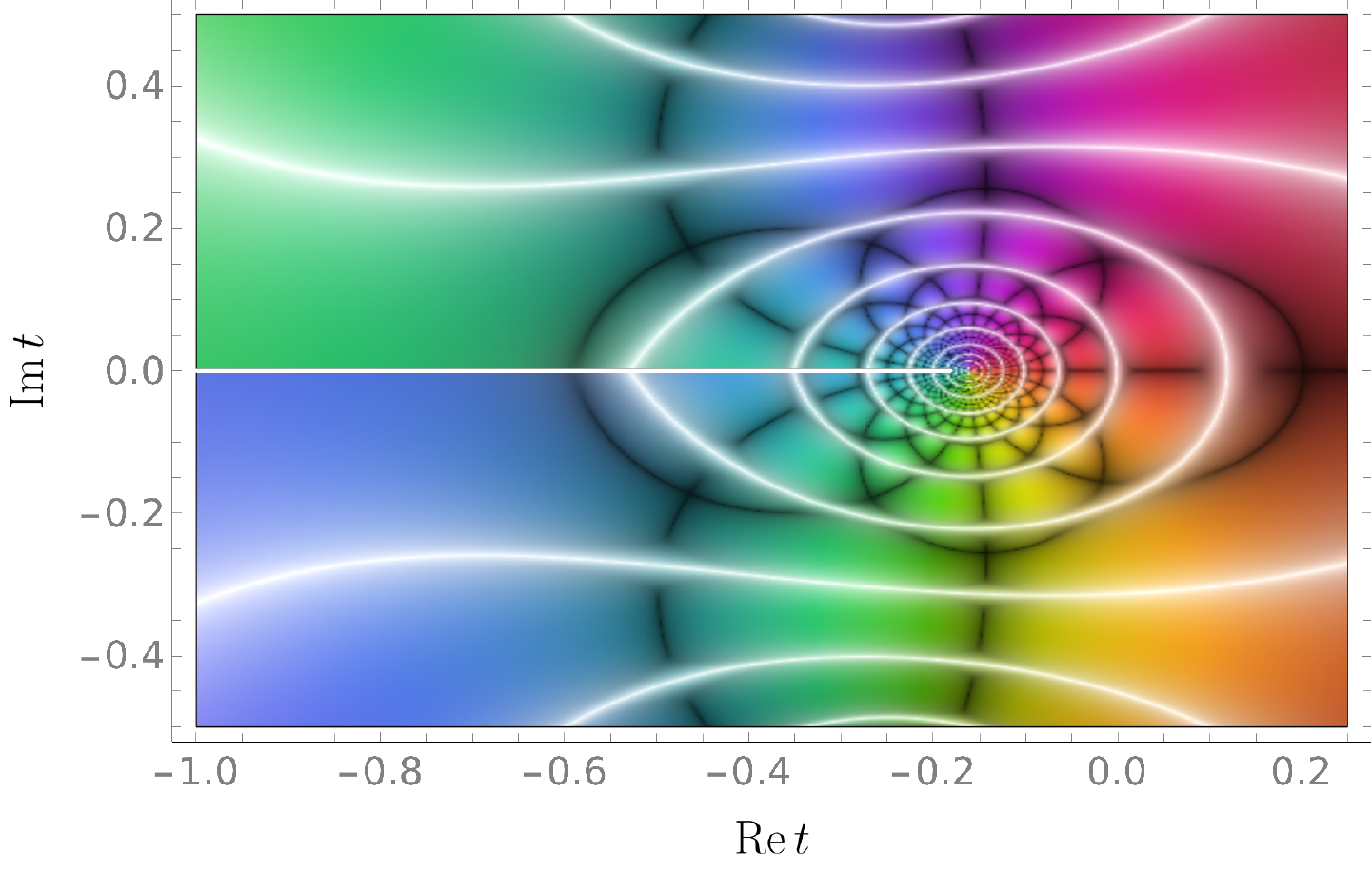}
    \caption{
        Hypergeometric $_{2}F_1$ approximant to $\operatorname{Borel}[B](t)$.
        The built-in branch cut along the negative real axis is evident as a discontinuity in the colouring.
        At this low order the estimate for the branch point is not very accurate, but this improves at higher order, cf.~\cref{fig:hypergeometric-cut}
    }
    \label{fig:hypergeometric}
\end{figure}

\begin{figure}[tbp]
    \centering
    \includegraphics[width=\linewidth]{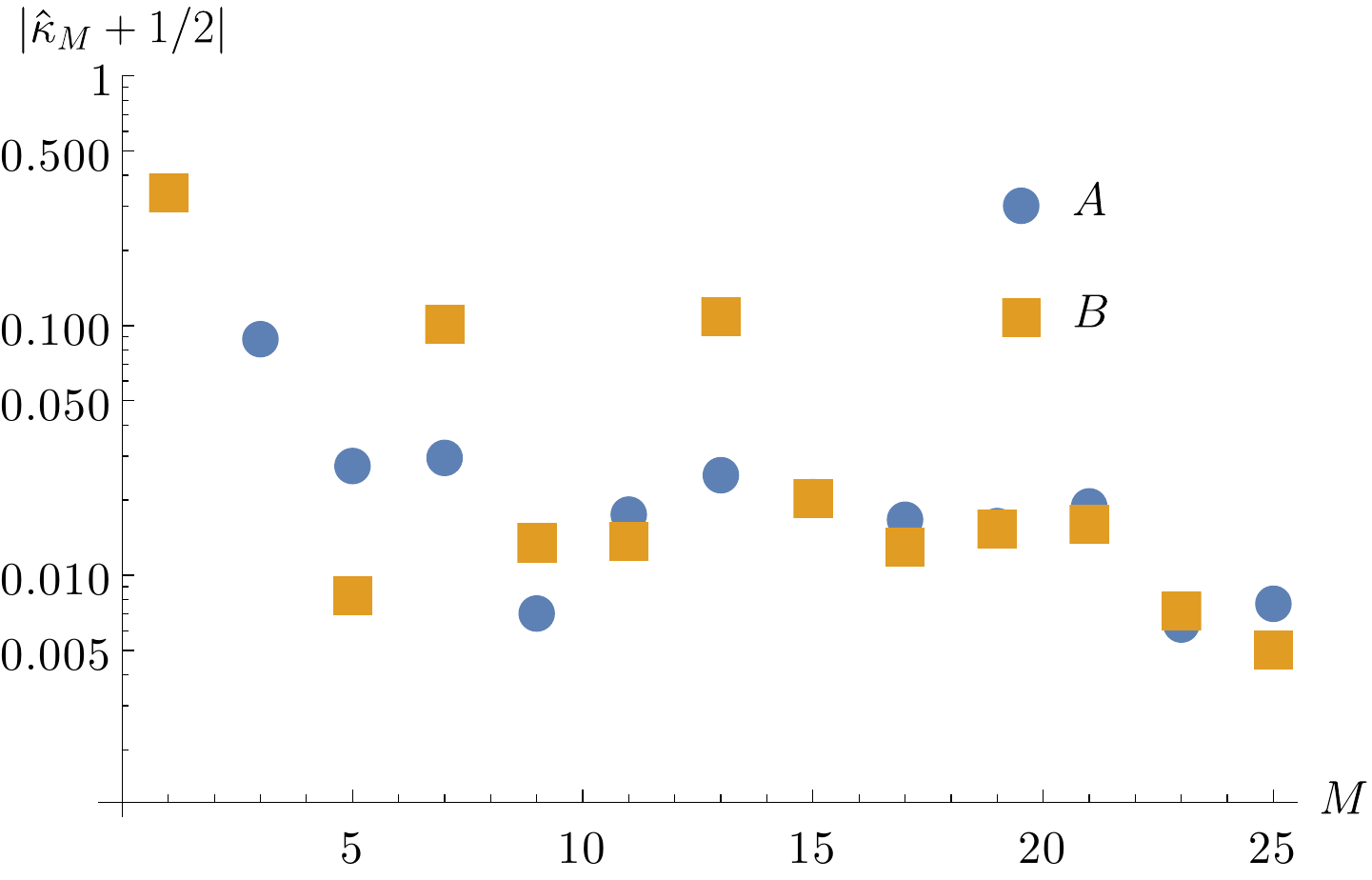}
    \caption{
        Estimates of the branch point using a hypergeometric approximant based on perturbative coefficients up to order $2M + 1$.
    }
    \label{fig:hypergeometric-cut}
\end{figure}

Instead of a Pad\'e approximant, we can use a hypergeometric approximant in the Borel plane~\cite{Mera:2018qte}.
With perturbative data up to order $N = 2M + 1$ a hypergeometric $_{M+1}F_M(\cdots, \cdots ; t/\hat{\kappa_M} )$ can be fitted;
it has built-in a branch cut at $\hat{\kappa}_M$.
\Cref{fig:hypergeometric} shows an example $_{2}F_1$ approximant for $\operatorname{Borel}[B](t)$, and \cref{fig:hypergeometric-cut} how $\hat{\kappa}_M$ converges to $\kappa = - \frac{1}{2}$.

We now return to the question of the sign of $\kappa$.
While having exponentially \emph{large} terms seems to be against the spirit of perturbation theory, in a purely formal treatment there is no ``wrong sign'' for $\kappa$, which may even be complex.
An instructive example (discussed in detail in Ref.~\cite[Sec.~2]{Marino:2012zq}) is the Airy functions, which have expansions
\begin{equation}
        2 \Ai(z), \Bi(z)
        \sim \\
        \frac{z^{-1/4}}{\sqrt{\pi}} e^{\mp \frac{2}{3} z^{3/2} }
        \left(1 + \ordo{z^{-3/2}} \right)
\end{equation}
as $z \to + \infty$ along the real axis.
The exponentially large $\Bi$ is a valid solution to the Airy equation; it just does not match the boundary condition $f(+\infty) = 0$.
As $z \to -\infty$ both $\Ai$ and $\Bi$ become oscillatory, corresponding to an imaginary $\kappa$.
This is the eponymous phenomenon first studied by Stokes~\cite{Stokes1851,Stokes1864} in precisely the context of the Airy functions.
(For a physical example with imaginary $\kappa$, see Ref.~\cite{Dunne:2021acr}.)

For LAD the initial acceleration is to be specified, while for $\LL\infty$ it is determined by the initial momentum and $\CA, \CB$ at $\delta_0 = \tau_0 \cE p_0^\LCp$.
Only the ODE:s~\cref{eq:fixed-point} need to hold for a solution of $\LL\infty$ to be a solution to LAD;
hence the choice of initial condition for the ODE:s~\cref{eq:fixed-point} determines which, among all solutions of LAD with a given inital momentum, is picked out by $\LL\infty$.

\begin{figure}[tbp]
    \centering
    \subfloat[]{\includegraphics[width=0.48\textwidth]{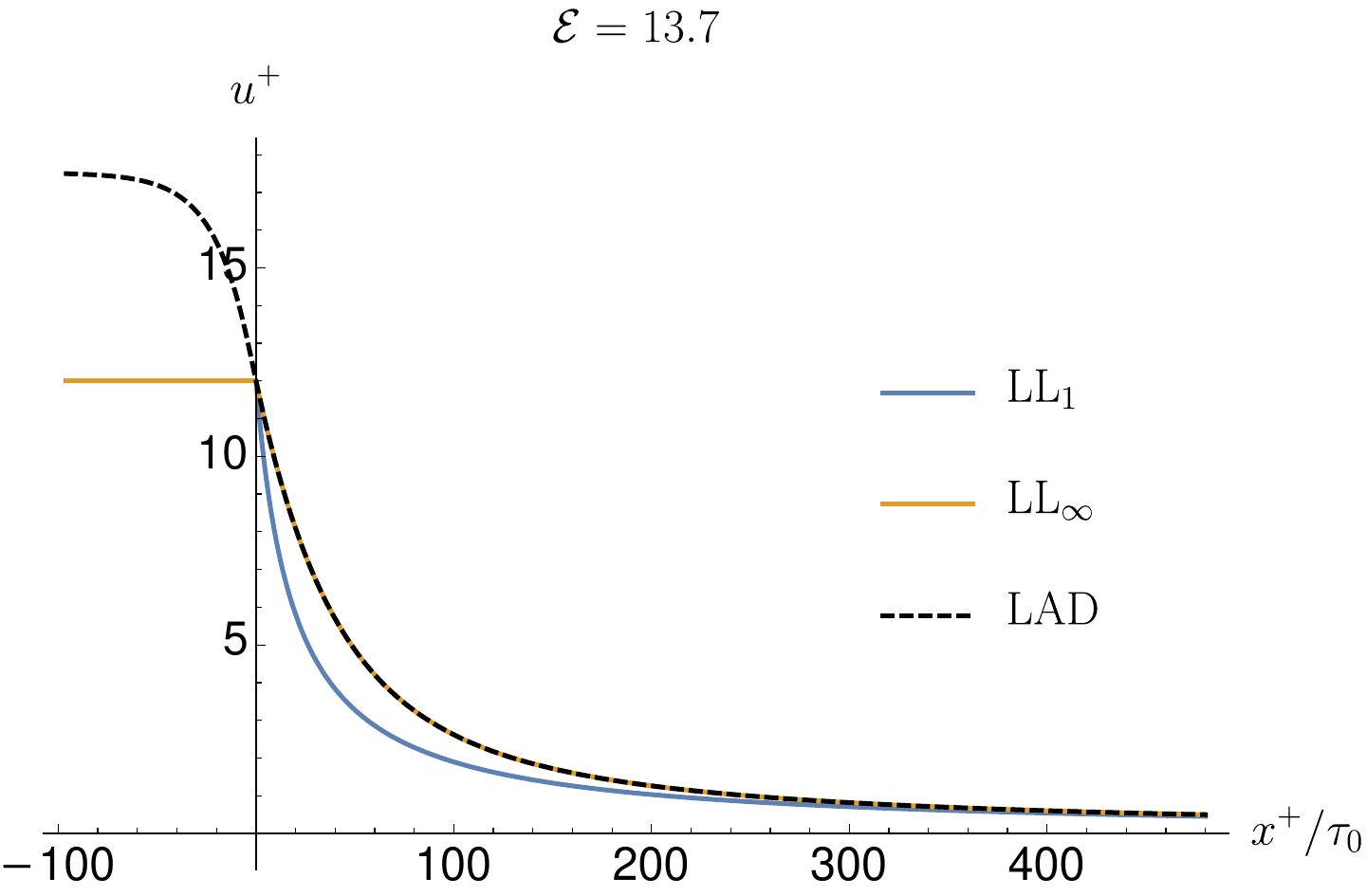}}\\
    \subfloat[]{\includegraphics[width=0.48\textwidth]{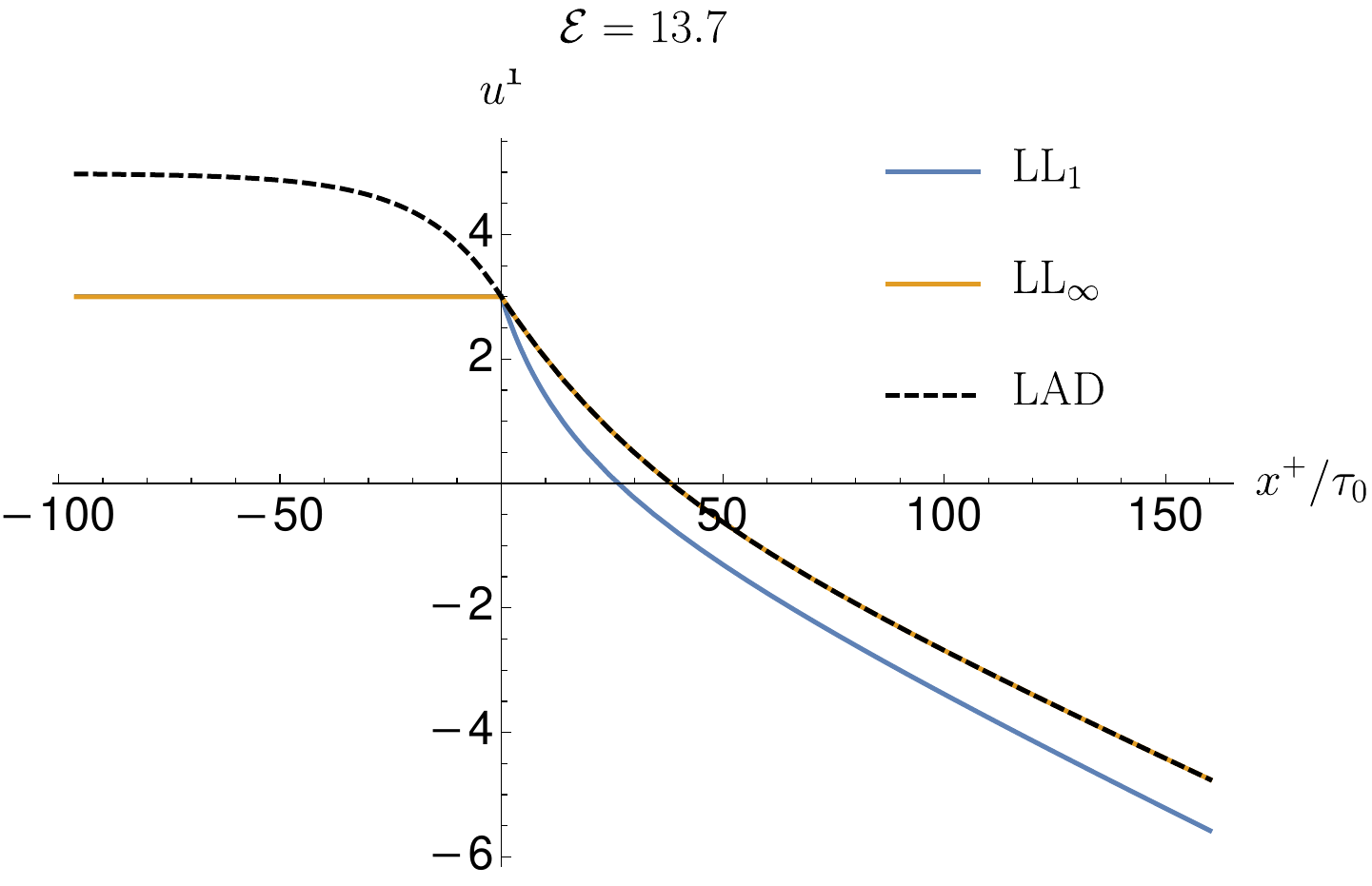}}
    \caption{
        Longitudinal (a) and transverse (b) momentum components across a step.
        $\LL\infty$ is seen to agree with the physical solution of LAD after the step, while the latter exhibits pre-acceleration.
        The pre-acceleration occurs over a few $\tau_0$ worth of proper time $\sim x^\LCp/p^\LCp$
    }
    \label{fig:LAD-LLinf-step}
\end{figure}

By dotting $n^\mu$ into and squaring~\cref{eq:LL-infinity}, respectively, we find that $\LL\infty$ implies
\begin{equation}
    \dot{p}^\LCp = - \tau_0 p^\LCp \CB(\delta) \delta^2
\end{equation}
and
\begin{equation}
    \label{eq:pdot-sq}
    \tau_0^2 \dot{p}^2 = -\CA(\delta)^2 \delta^2 - \CB(\delta)^2 \delta^4
    \,
    .
\end{equation}
With the Lorentz initial condition~\cref{eq:A-B-ic} the resummed perturbative $\CA_\text{pert}, \CB_\text{pert}$ are positive and approach $1$ smoothly as $\delta \to 0$.
This means that $\delta \to 0$ and thence $\dot{p}^2 \to 0$ as $\tau \to \infty$.
The solution of $\LL\infty$ is therefore the \emph{physical, non-runaway, solution} of LAD, shown in \cref{fig:LAD-LLinf-step}.
In other words, $\LL\infty$ with the Lorentz initial condition~\cref{eq:A-B-ic} determines the \emph{critical acceleration} (a concept first introduced in Ref.~\cite{Spohn:1999uf})
\begin{equation}
    \label{eq:crit-acc}
    \dot{p}^\mu_\text{crit}
    :=
    \CA_\text{pert}(\delta_0) f^{\mu}{}_{\nu} p^\nu_{0}
    + \CB_\text{pert}(\delta_0) (Pf^2)^{\mu}{}_{\nu} p^\nu_{0}
\end{equation}
that, for a given field strength and initial momentum, leads to the physical solution of LAD;

As the purely perturbative solution of~\cref{eq:fixed-point} leads to the physical solution to LAD, the remaining, non-perturbative, solutions must lead to the runaways.
We cannot find non-perturbative solutions with an initial condition at $\delta = 0$, but we can equally well set the initial condition at $\delta_0 = \tau_0 \cE p^\LCp_0$.
Again using the Airy functions to illustrate, with the boundary condition $f(+\infty) = 0$ we discard $\operatorname{Bi}$, but setting a condition at finite argument retains it.
The solution of~\cref{eq:fixed-point} satisfying
\begin{subequations}
    \label{eq:non-pert-ic}
    \begin{align}
        \CA(\tau_0 \cE p^\LCp_0) & = \CA_\text{pert}(\tau_0 \cE p^\LCp_0) \\
        \CB(\tau_0 \cE p^\LCp_0) & = \CB_\text{pert}(\tau_0 \cE p^\LCp_0) + \frac{\tau_0}{\delta^2_0} \varepsilon
        \,
        .
    \end{align}
\end{subequations}
will give us a solution to LAD with an initial longitudinal acceleration differing from the critical by $\varepsilon$;
we expect this solution to be a runaway.

The general solution~\cref{eq:gen-sol} with $c_1 = 0, c_2 = -\tau_0 \varepsilon e^{-1/2\delta_0^2} $ verifies the initial condition~\cref{eq:non-pert-ic}.
This is only the leading term at first non-perturbative order, but it will be sufficient.
This gives us for the longitudinal acceleration
\begin{equation}
    \begin{split}
        \frac{\ud p^\LCp}{\ud x^\LCp} = - \frac{\delta^2}{\tau_0} \CB(\delta)
        = &
        -\frac{\delta^2}{\tau_0} \Big[
            1 - 6 \delta^2 + \ldots \\
          &    + \frac{\varepsilon}{\delta^2} \exp \big(
              \frac{1}{2\delta^2} - \frac{1}{2\delta_0^2}
          \big)
      \Big]
      \,
      .
    \end{split}
\end{equation}
After a short time $\delta(x^\LCp) \approx \delta_0 + x^\LCp \cE \tau_0 \frac{\ud p^\LCp}{\ud x^\LCp}$ and using this to expand we have
\begin{equation}
    \label{eq:runaway-acc}
    \frac{\ud p^\LCp}{\ud x^\LCp} = - \frac{\delta^2_0}{\tau_0} \Big(
        \CB_\text{pert}(\delta_0)
        +
        \frac{\tau_0 \varepsilon}{\delta^2_0} e^\frac{x^\LCp}{\tau_0 p^\LCp_0}
        + \ldots
    \Big)
\end{equation}
omitting some inessential terms.
Clearly the second term inside the brackets is a runaway over a proper time $\tau_0$, and it is only seen because we included non-perturbative terms in $\CB$.

It is clear from~\cref{eq:pdot-sq} that if $\CB(\delta_0) > 0$ and $\CB$ remains positive as $\delta$ decreases we have a runaway in the $-\hat{z}$ direction.
Likewise if $\CB(\delta_0) < 0$ and keeps its sign as $\delta$ increases we have a runaway in the $+\hat{z}$ direction.
It is less obvious what happens if $\CB$ should cross $0$.
This is relevant because we can imagine perturbing $\dot{p}^\mu_\text{crit}$ in the $+\hat{z}$ direction, seeding a runaway instability.
Initially, then, $p^\LCp$ is decreasing, but after some time the instability begins to dominate and $p^\LCp$ increases.
If the dynamics are indeed described by $\LL\infty$, $\CB$ must change sign when this happens.

\begin{figure}[tb]
    \centering
    \subfloat[\label{fig:branches:a} ]{
        \includegraphics[width=\linewidth]{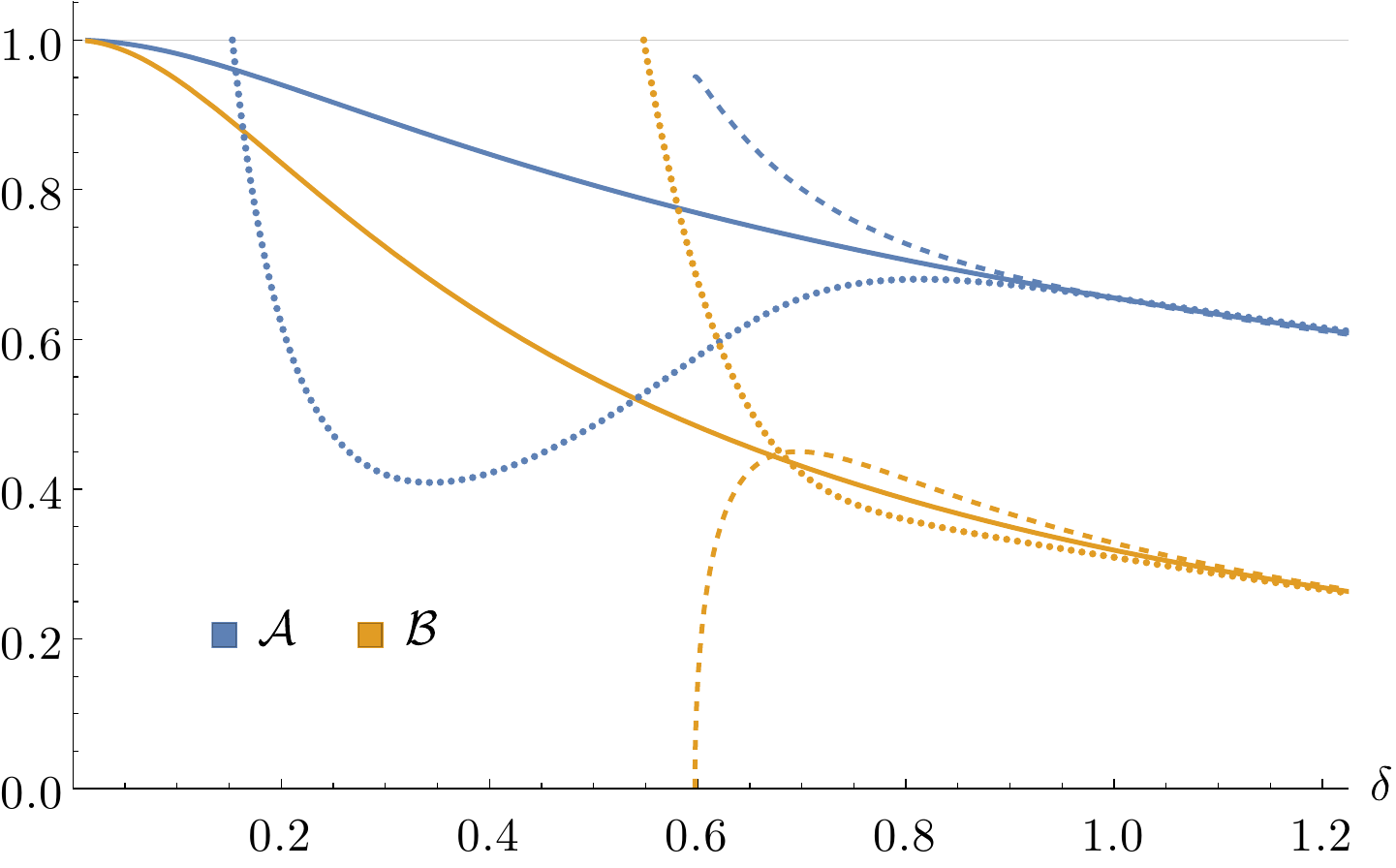}
    }\\
    \subfloat[\label{fig:branches:b} ]{
        \includegraphics[width=\linewidth]{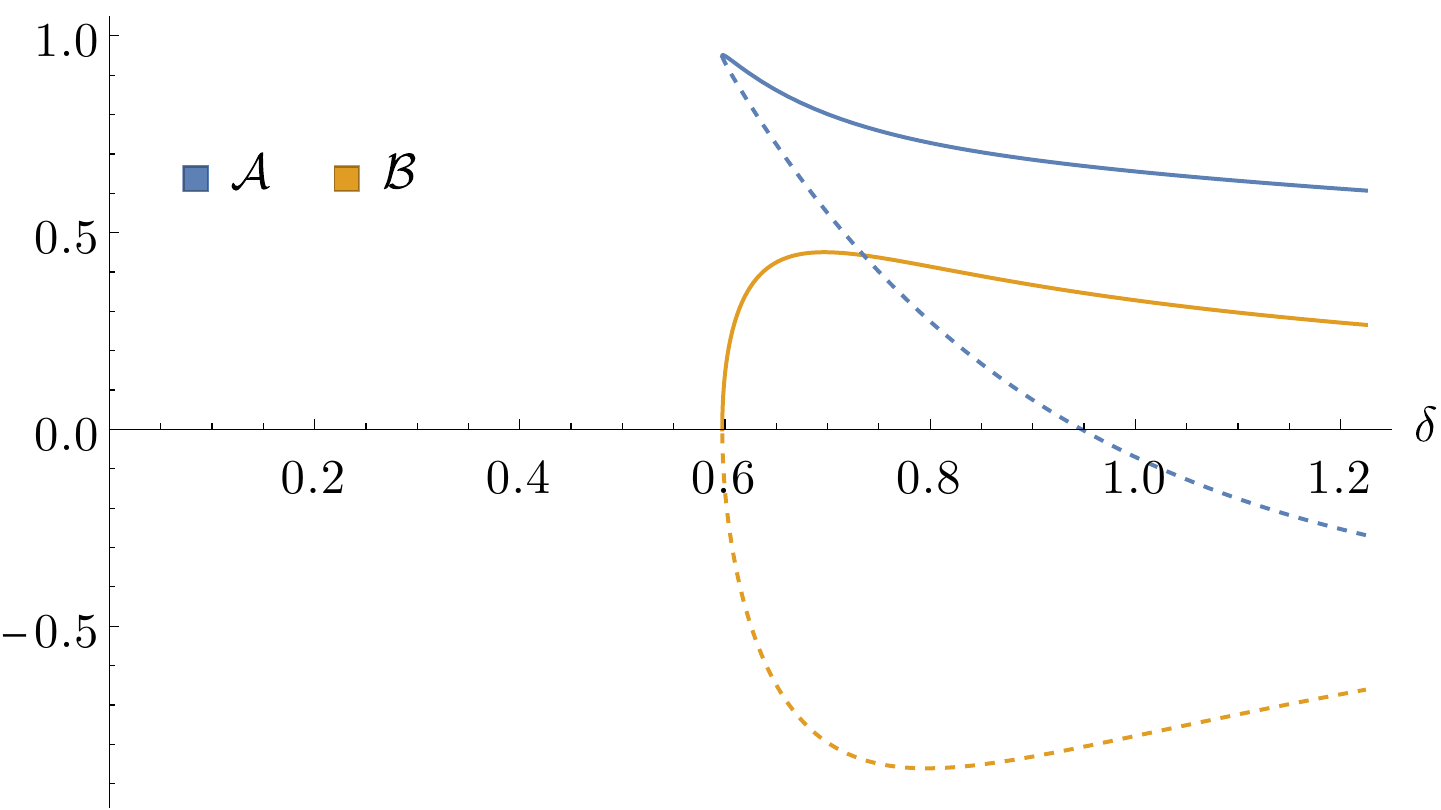}
    }
    \caption{
        (a) Three solutions to the ODE:s~\cref{eq:fixed-point}.
        The solid curves represent the resummed perturbative solution smoothly approaching $1$ as $\delta \to 0$;
        the dotted and dashed curves represent non-perturbative solutions: the former blowing up exponentially at small $\delta$,
        the latter approaching the singular point $\CB = 0$.
        (b) At the singular point $\CB = 0$ the solution is non-unique.
        There is an ``upper'' (solid) and a ``lower'' branch (dashed), characterised by $\CB \approx \pm |\CA(\tilde{\delta})| \sqrt{1/\tilde{\delta}^2 - 1/\delta^2} $, respectively, near the singularity.
    }
    \label{fig:branches}
\end{figure}

\Cref{fig:branches:a} shows the perturbative solution, as well as two solutions with initial conditions
\begin{equation}
    \CA_\pm(1) = \CA_\text{pert}(1) \qquad
    \CB_\pm(1) = (1 \pm \varepsilon)\CB_\text{pert}(1)
    \, .
\end{equation}
Of these, $\CB_-$ runs away to $+\infty$ as $\delta$ decreases, and so gives the $-\hat{z}$ runaway,
while instead $\CB_+$ approaches $0$ at a finite argument $\tilde{\delta}$.
Now, $\CB = 0$ is a singular point of~\cref{eq:fixed-point}, but to leading order around it the system reads
\begin{equation}
    \left\{ \begin{aligned}
        0 & = 1 - \CA \\
        \delta^3 \CB \frac{\ud \CB}{\ud \delta} & = \CA^2
    \end{aligned}
    \right.
    \, ,
\end{equation}
with the \emph{two} solutions
\begin{equation}
    \CB \approx \pm \big| A(\tilde{\delta}) \big| \sqrt{1/\tilde{\delta}^2 - 1/\delta^2}
    \, .
\end{equation}
These ``branches'' are both shown in~\cref{fig:branches:b}.
What happens, then, for the $+\hat{z}$ runaway is that we start out on the upper branch, but as $\dot{p}^\LCp$ becomes $0$ and the runaway instability begins to dominate,
the dynamics continue to be described by $\LL\infty$, but now on the lower branch.

\begin{figure}[tb]
    \centering
    \includegraphics[width=\linewidth]{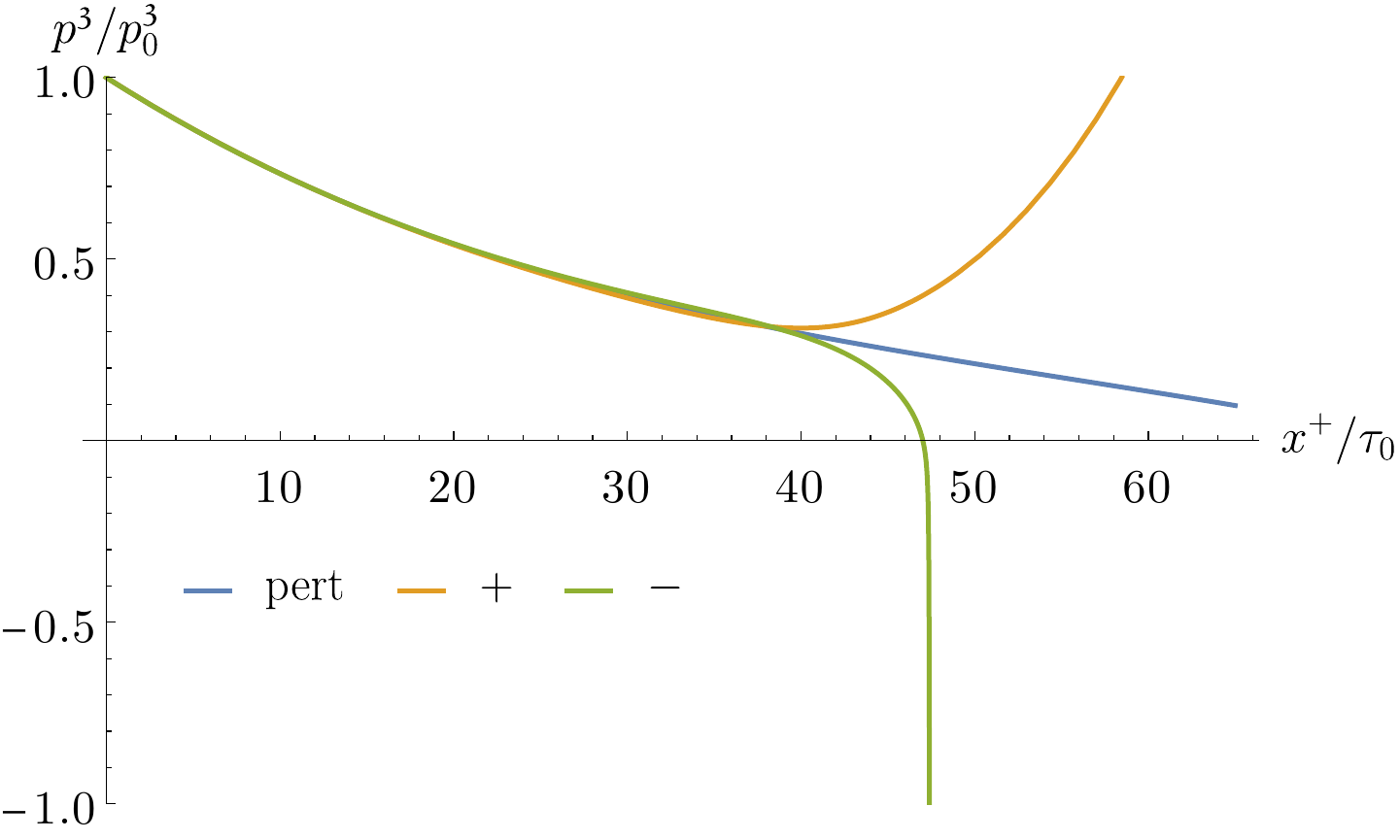}
    \caption{
        $z$-component of momentum for three solutions of LAD with \emph{initial} accelerations given by $\LL\infty$, using
        either $\CA_\text{pert}, \CB_\text{pert}$ (blue) or $\CA_\pm, \CB_\pm$ (gold, green).
        The two latter are runaway solutions in the $\mp \hat{z}$ direction, respectively; while the first is the physical solution.
    }
    \label{fig:lad-runaways}
\end{figure}

\begin{figure}[tb]
    \centering
    \includegraphics[width=\linewidth]{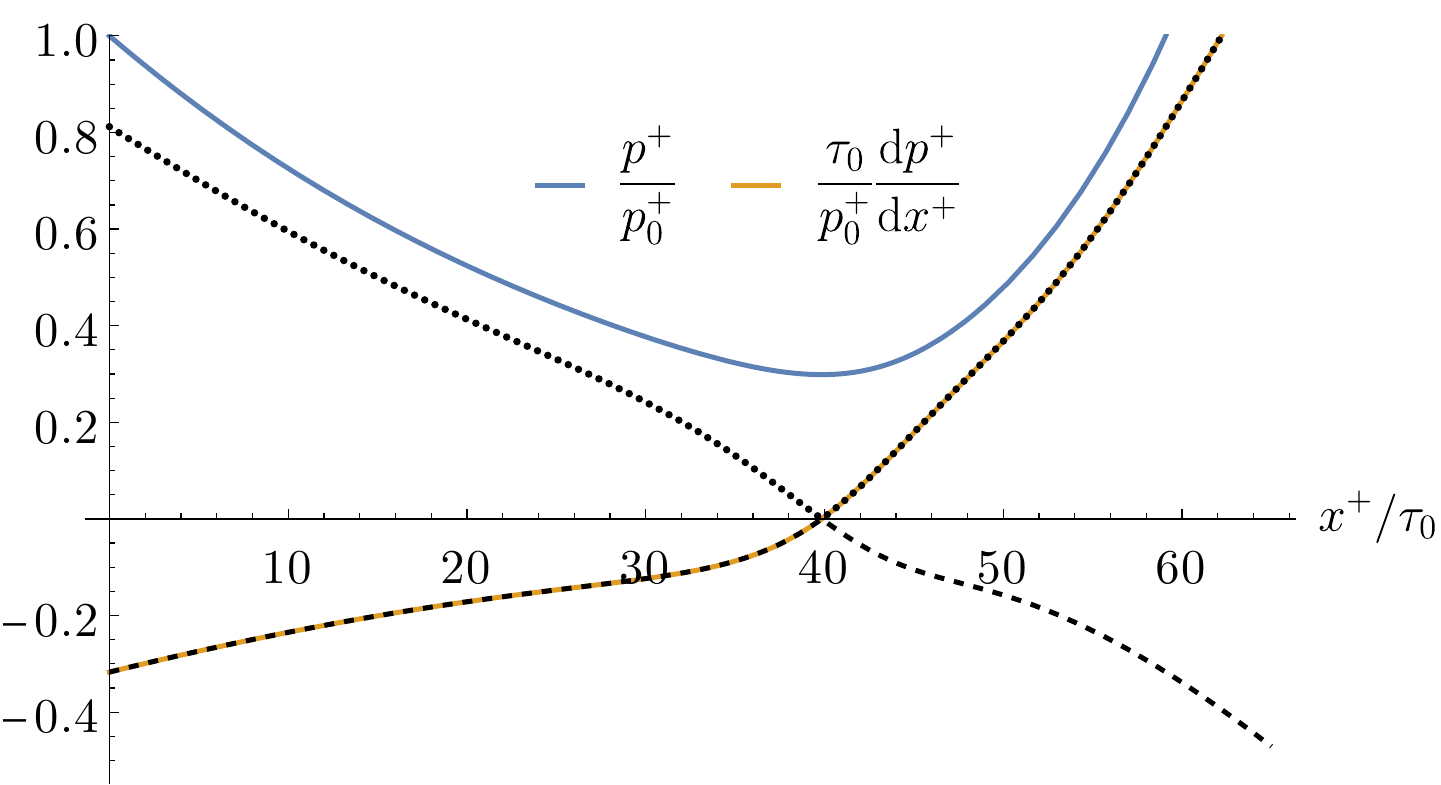}
    \caption{
        Longitudinal momentum $p^\LCp$ (blue) and acceleration (gold) for a runaway solution  of LAD in the $+\hat{z}$ direction.
        The dashed and dotted curves indicate the $\LL\infty$ acceleration $\propto \CB(\tau_0 \cE p^\LCp)$, using the upper and lower branches for $\CB$, respectively; cf. \cref{fig:branches:b}.
        It is seen that as the longitudinal acceleration becomes zero it switches between the two branches.
    }
    \label{fig:lad-runaway-acc}
\end{figure}

To verify this we solve LAD numerically with the initial accelerations as implied by $\LL\infty$, using $\CA_\pm, \CB_\pm$.
Plotting $p^z$ in \cref{fig:lad-runaways} we find that the respective initial conditions indeed lead to the runaways we expect from \cref{fig:branches:a}
We also observe the switching between branches by plotting $p^\LCp$ and its derivative for the $+\hat{z}$ runaway and comparing with $\LL\infty$, see \cref{fig:lad-runaway-acc}.
We stress that in \cref{fig:lad-runaways,fig:lad-runaway-acc} we have solved LAD \emph{forward} in time:
because the initial acceleration is either the critical or very close to it, the instability remains suppressed for several $\tau_0$ worth of proper time.

We end this Section by noting that the form of~\cref{eq:LL-infinity} is fully determined by there being only two possible tensor structures and one scalar invariant ($\delta$) in the CCF geometry.
Another highly restricted geometry is a circularly polarised monochromatic plane wave, and it is possible to derive equations similar to~\cref{eq:fixed-point}, and hence $\LL\infty$ also in that case.
It can be studied with the Borel plane methods we have applied to the CCF in this Section;
as the details are very similar, we defer them to \cref{app:monochromatic}.

In either case, we obtain that
reduction of order eliminates non-perturbative terms on the level of the \emph{equation of motion} when a physical boundary condition -- matching to the Lorentz force at vanishing field intensity -- is imposed.
We therefore now turn to how non-perturbative pre-accelerating and runaway solutions arise on the level of \emph{solutions} to LAD.

\section{LAD and \texorpdfstring{$\LL\infty$}{LL∞} in a crossed step field}
\label{sec:step}

We will now consider LAD and $\LL\infty$ in a field with a step profile, i.e.
\begin{equation}
    a'_\mu = \mathcal{E} \theta(x^\LCp) \epsilon_\mu
    \,
    .
\end{equation}
That the field is off for an interval of time will allow an unambiguous identification of pre-acceleration.

For LAD we are faced with the problem of matching solutions before and after the step.
The integro-differential form of LAD~\cite{Rohrlich1961,Plass:1961zz,PhysRevE.88.033203}, however, shows that the acceleration is continuous across a step, and so
we should use the critical acceleration~\cref{eq:crit-acc}

\subsection{ Exact Solution to Free LAD }

Before the step, with the field turned off, all the equations of motion can be solved exactly.
For $\LL1$ and $\LL\infty$ the solution is just uniform motion, while for LAD we make an Ansatz in terms of proper time $\tau$ and the rapidity $\zeta$,
\begin{equation}
    \label{eq:LAD-Ansatz}
    p^\mu(\tau) = \cosh(\zeta(\tau)) p^\mu_0 + \sinh(\zeta(\tau)) \frac{\dot{p}_0^\mu}{\sqrt{-\dot{p}_0^2}}
    \,
    ,
\end{equation}
where the subscript $0$ indicates values at $\tau = 0$.
LAD then implies an initial-value problem for $\zeta$,
\begin{equation}
    \label{eq:zeta-ivp}
    \tau_0 \ddot{\zeta} = \dot{\zeta}
    \qquad
    \zeta(0) = 0
    \quad
    \dot{\zeta}(0) = \sqrt{-\dot{p}_0^2}
    \,
    ,
\end{equation}
with solution
\begin{equation}
    \label{eq:zeta-sol}
    \zeta = \tau_0 \sqrt{-\dot{p}_0^2} \left(  e^{\tau/\tau_0} - 1 \right)
    \,
    .
\end{equation}
We see that the pre-step solution is pre-accelerating unless $\dot{p}_0^\mu = 0$.
Viewed forwards in time this solution generalises the well-known non-relativistic runaway in that the exponential runaway is in the rapidity, rather than in the velocity.
To the best of our knowledge, the covariant solution matched to the initial conditions, \cref{eq:LAD-Ansatz,eq:zeta-sol}, has not previously appeared in the literature%
~\footnote{%
    An expression involving $\sinh$ appears in Ref.~\cite{Plass:1961zz} but the approach there is not explicitly covariant and only seeks to prove that the non-runaway solution is uniform motion.
}.

The solution has the form of a transseries in $\tau_0$ with, expanding the hyperbolic functions, non-perturbative instanton terms of all orders.
For $\tau > 0$ these become \emph{large} as $\tau_0 \to 0$, corresponding to faster runaways;
for $\tau < 0$ they become  \emph{small} in this limit, corresponding to the pre-acceleration occurring in a ``boundary layer'' of width $\approx 1/\tau_0$.

Note, though, that the solution is analytic in the proper time $\tau$:
the pre-factor of each $e^{\ell \tau/\tau_0}$ term is some power series in $\tau_0$, cf.~\cref{eq:crit-acc}.
In fact $\tau$ only appears as $\tau/\tau_0$ and after a change of variables $(\tau, \tau_0) \to (\tilde{\tau}, \tau_0) = (\tau/\tau_0, \tau_0)$
the solution is analytic in both variables.
This can be traced to that in free LAD, or equivalently~\cref{eq:zeta-ivp} the only scale is $\tau_0$, which can be eliminated by rescaling.
There is then no coupling in which to do perturbation theory, but the equation can be solved as a power series in rescaled time;
$\tau_0$ reenters when substituting for the initial condition~\cref{eq:crit-acc}.

This is a simple demonstration that the character of a transseries in \emph{two} variables can change dramatically with a non-linear change of variables.
Such non-linear transformations can in effect perform partial resummations in one of the variables,
a point previously discussed in the contexts of a unitary matrix model~\cite{Ahmed:2017lhl} and radiation reaction~\cite{Torgrimsson:2021zob}.
As a consequence there can be subtleties in how (e.g., in which order) limits are taken; we will return to this shortly.
(See also Refs.~\cite{Podszus:2018hnz,Ilderton:2019kqp} for another example in strong-field physics where the manner of taking a limit matters.)

The above discussion has been in terms of proper time only while the rest of this paper uses lightfront time.
We therefore conclude this subsection with a short discussion of the solution of free LAD in lightfront parametrisation.
In lightfront time the equation for the rapidity retains factors of $\cosh \zeta, \sinh \zeta$ and cannot be solved analytically.
Alternatively we can obtain the lightfront time by quadrature,
\begin{equation}
    \label{eq:xplus-implicit}
    x^\LCp(\tau) = \int_0^\tau \ud \sigma \,
        p^\LCp(\sigma)
    \,
    .
\end{equation}
While this integral does have an analytic expression in terms of $\Ei(\cdot)$, it gives only an implicit relation for $\tau(x^\LCp)$.
It can, though, be expanded to NLO $\tau/\tau_0$ to find
\begin{equation}
    \tau/\tau_0 = p^\LCp_0 x^\LCp/\tau_0 - \frac{\tau_0}{2} \dot{p}_0^\LCp (x^\LCp/\tau_0)^2
    + \ordo{(x^\LCp/\tau_0)^3}
    \,
    .
\end{equation}
Inserting this back into~\cref{eq:LAD-Ansatz,eq:zeta-sol}
yields another example of a non-linear transformation strongly modifying the two-variable transseries structure.

\subsection{ Transseries Solution of LAD }

We now come to the transseries structure of solutions to LAD in a constant crossed field.
\cemph{
    This was briefly studied in Ref.~\cite{Torgrimsson:2021zob} and we are mainly concerned with working out some implications of the results therein.
    Using notation slightly different from Ref.~\cite{Torgrimsson:2021zob}, we can formulate LAD in a CCF as
}
\begin{subequations}
    \label{eq:torgrimsson}
    \begin{align}
        g' & = \delta \left[ \partial_u ( g g' ) + g^2 P \right] \\
        h' & = 1 + \delta \left[ \partial_u ( g h' ) + g h P \right]
        \label{eq:torgrimsson-b}
        \\
        P & = (g')^2  - (h')^2 + 2 g' \partial_u \frac{1 + h^2  - g^2}{2 g} \notag
        \,
        .
    \end{align}
\end{subequations}
Here $g$ and $h$ are normalised longitudinal and transverse components respectively%
~\footnote{\cemph{%
    It is not necessary to normalise to specifically $p_0$; any on-shell momentum will do.
    This choice is the most convenient for the present purpose, though.
    Other choices imply initial conditions other than~\cref{eq:instanton-ic-acc,eq:instanton-ic-mom}.
}
},
\begin{subequations}
    \begin{align}
        g & := p^\LCp/p_0^\LCp \\
        h & := g p^\mathfrak{1}_0 - p^\mathfrak{1}
        \,
        ,
    \end{align}
\end{subequations}
the prime is a derivative with respect to a normalised lightfront time $u := \cE x^\LCp$, and $\delta^2 = \tau_0^2 p_0 f^2 p_0$, i.e., we drop the subscript on $\delta_0$ from the previous Section.

Ref.~\cite{Torgrimsson:2021zob} solves these equations iteratively by noting that if $g, h$ have series expansions in $\delta$,
with the coefficients being functions of time,
\begin{equation}
    \label{eq:gh-ansatz}
    \begin{Bmatrix}
        g \\ h
    \end{Bmatrix}
    \sim
    \sum_n \delta^n
    \begin{Bmatrix}
        g_n(u) \\ h_n(u)
    \end{Bmatrix}
\end{equation}
the order $n$ terms of the RHS are determined by terms of strictly lower order, so $g_n, h_n$ can be found iteratively by simple integration.
The zeroth order starting point is $g_0 = 1, h_0 = u$, corresponding to the Lorentz force.
The coefficients are polynomials in $u$, with the first few being as follows:
\begin{widetext}
    \begin{subequations}
    \label{eq:gh-sol}
        \label{eq:torgrimsson-sol-perturb}
        \begin{align}
            g(u) & = 1 - u \delta + u^2 \delta^2 + (6u - u^3) \delta^3 + (-18 u^2 + u^4) \delta^4 + \ordo{\delta^5} \\
            h(u) & = u - \frac{1}{2} u^2 \delta
            + \left(-2u + \frac{u^3}{2}\right)  \delta^2
            + \left(6u^2 - \frac{u^4}{2}\right) \delta^3
            + \left(20 u - \frac{41 u^3}{3} + \frac{u^5}{2u} \right) \delta^4
            + \ordo{\delta^5}
            \,
            .
        \end{align}
    \end{subequations}
\end{widetext}
Notably $g'(0), h'(0)$ have precisely the same perturbative expansion as one would find using $\LL\infty$ for $\dot{p}^\mu_0$.

The solution~\cref{eq:torgrimsson-sol-perturb} also illustrates the care needed in taking limits in formal, divergent expansion.
At each order in $\delta$ the leading behaviour in $u$ of $g$ is $(-u\delta)^n$, the series has a finite radius of convergence, and can be resummed into $1/(1 + u \delta)$, which is the exact solution of $\LL1$ \cite{Heintzmann:1972mn,PiazzaExact}.
This has a single pole in the complex plane
~\footnote{
    The pole indicates that there is a minimum lightfront time in the past at which the particle was at the speed of light,
    cf.~Refs.~\cite{Tomaras:2000ag,Woodard:2001hi,Ekman:2021vwg}
}
and its Borel transform ($e^{-\delta t}$) is everywhere analytic.
For any fixed $u$ the linear term $\sim n! u$ will always win over $u^n$, though, meaning that the $u \to \infty$ limit must be taken \emph{inside} the sum in~\cref{eq:gh-ansatz}.

If this iterative method is applied to free LAD (which corresponds to striking the constant term on the RHS of~\cref{eq:torgrimsson-b}), only the ``trivial'' solution of uniform motion is found.
It is to be expected that solutions are lost as the method is only sensitive to initial conditions for the momentum, not the acceleration.
In either case, the generated perturbative solution is the physical solution (but must be resummed), and we must introduce non-perturbative transseries terms to capture pre-acceleration and runaways.

To find all solutions, including pre-accelerating and runaway solutions, instead of a simple series in $\delta$, then, we should use a transseries Ansatz~\cite{Torgrimsson:2021zob},
\begin{equation}
    \label{eq:trans-series-std}
    g(u)  \sim \sum_{n, \ell} \delta^n e^{\ell u/\delta} g_{n,\ell}(u)
    \,
    .
\end{equation}
We will refer to terms with $\ell \ge 1$ as \emph{instanton} terms by analogy with quantum theory~\cite{Lipatov:1977hj}, even though their origin is different.
Note again that the coefficients are functions of time -- as in the previous subsection this is an expansion in two variables.
The operator $\sim \delta \frac{\ud^2}{\ud u^2}$ on the RHS of~\cref{eq:torgrimsson} lowers by $1$ the degree in $\delta$ of any term with $\ell \ge 1$.
Thus we no longer have that $g'_{n, \ell}$ is determined by simply integrating lower-order coefficients, but rather by coupled first-order ODE:s.

The expansion~\cref{eq:trans-series-std} is also lacking in that the initial conditions for the $g_{n,\ell}$ are grossly underdetermined, as we only have
\begin{widetext}
 \begin{equation}
    \label{eq:instanton-ic-mom}
        g(0)  = 1 \sim \sum_n \delta^n \sum_\ell g_{n, \ell}(0)
        \quad \text{and} \quad
        h(0)  = 0 \sim \sum_n \delta^n \sum_\ell h_{n, \ell}(0)
\end{equation}
\end{widetext}

and similar initial conditions for the acceleration,
\begin{equation}
    \label{eq:instanton-ic-acc}
    g'(0) \sim \sum_n \delta^{n-1} \sum_\ell g_{n-1, \ell}'(0) + \ell g_{n, \ell}(0)
    \,
    .
\end{equation}

Since the zeroth and first derivatives of $g, h$ at $0$ determine all higher derivatives at $0$ through LAD~\cref{eq:torgrimsson}, there is in principle an infinite hierarchy of constraints resolving the underdetermination.
However, each rung of the ladder involves instanton terms of all orders, so we cannot proceed iteratively.
(The system is not ``triangular'', so to speak.)

We are thus unable to iteratively determine fully self-consistently the precise transseries form of a specified runaway or pre-accelerating solution.
We can however truncate the system to one-instanton terms and assume that their initial amplitudes are $\ordo{\varepsilon}$, which will be accurate to $\ordo{\varepsilon^2 e^{2u/\delta}}$.
To make contact with the preceding section we will look for a solution such that
\begin{subequations}
    \begin{align}
        g'(0) & = +\frac{\varepsilon}{\delta} - \delta \CB_\text{pert}(\delta) \\
        h'(0) & = \CA_\text{pert}(\delta)
        \,
        .
    \end{align}
\end{subequations}
This again corresponds to a runaway with initial acceleration $\varepsilon$ different from the critical.
(This is for concreteness only; the ODE:s for the instanton coefficients are linear and other initial conditions pose no greater problems.)

At $n = 0, \ell = 0$, we have $ g_{0,0} = 1 + \varepsilon, h_{0,0} = u$, keeping an integration constant that was implicitly dropped in the perturbative solution in order to account for instantonic contributions to the initial momentum.
This implies order $\varepsilon$ corrections to the following perturbative terms, beginning with $g_{1,0} = -2\varepsilon - u, h_{1,0} = + \varepsilon - \frac{u^2}{2}(\varepsilon + 1)$.
The $n = 0, \ell = 1$ components~\cite{Torgrimsson:2021zob} verify
\begin{widetext}
    \begin{equation}
        \label{eq:instanton-ode}
        \frac{\ud}{\ud u}
        \begin{pmatrix}
            g_{0,1} \\
            h_{0,1}
        \end{pmatrix}
        =
        \begin{pmatrix}
            \tilde{g}_{1,0} + u & -2 \\
            1 + 2u^2 & \tilde{g}_{1,0} - 3u
        \end{pmatrix}
        \begin{pmatrix}
            g_{0, 1} \\
            h_{0, 1}
        \end{pmatrix}
        \implies
        \begin{pmatrix}
            g_{0,1}(u) \\
            h_{0,1}(u)
        \end{pmatrix}
        =
        -\varepsilon e^{u^2/2}
        \begin{pmatrix}
            \cos 2u \\
            u \cos 2u - \sin 2u
        \end{pmatrix}
        \,
        .
    \end{equation}
\end{widetext}
For any initial acceleration other than the critical the instanton coefficients $g_{0,1}, h_{0, 1}$ grow superexponentially, i.e.,
we have a runaway solution.

This procedure can in principle be iterated to any instanton order and any order in $\delta$, although expanding the RHS of~\cref{eq:torgrimsson} becomes progressively costlier.
At order $\delta^n e^{\ell u/\delta}$ the instanton coefficients take the form $ \varepsilon^\ell \operatorname{Re} \left[ P_{n,\ell} (u) e^{\ell (u^2/2 - 2 i u) } \right] $ for some complex polynomial $P_{n,\ell} $ of degree $n$.
We have calculated $P_{n,1}$ up to $n = 16$, for which the constant terms and leading coefficients grow factorially and exponentially, respectively.
Hence just like the perturbative series, the instanton series must also be resummed for small $u$, but are convergent when the limit $u \to \infty$ is taken inside the sum.
\cemph{
    We stress that this result, as well as~\cref{eq:instanton-ode,eq:gh-sol}, agrees with Ref.~\cite{Torgrimsson:2021zob}.
}

The Gaussian form can be understood as the instanton coefficients reconstructing the non-trivial dependence $\tau(x^\LCp)$.
For the free solution,
\begin{equation}
    \label{eq:xp-tau}
    \frac{\tau}{\tau_0} \approx \frac{x^\LCp}{\tau_0 p^\LCp_0} - \frac{(x^\LCp)^2 \dot{p}^\LCp_0}{\tau_0 (p_0^\LCp)^3}
    = \frac{u}{\delta} + \frac{u^2}{2}
    + \ordo{u^3 \delta}
\end{equation}
when the initial acceleration is (close to) the critical.
Because the quadratic term is independent of $\delta$ it appears separately at each order and the modification to the exponent can be read off directly.
The next term in the exponent, going like $u^3 \delta$, cannot be identified at a single order in $\delta$, but would appear in an explicit resummation.

\section{Conclusions}
\label{sec:concs}

We have used the Lorentz-Abraham-Dirac (LAD) equation for radiation reaction (RR) as a ``laboratory'' setting in which to
probe non-perturbative physics using transseries methods.
Our choice of LAD for this purpose is motivated both by a large current interest in radiation reaction~%
\cite{Burton:2014wsa,Blackburn:2019rfv,Gonoskov:2021hwf,Damour:2020tta,DiVecchia:2021bdo,Herrmann:2021tct,Bjerrum-Bohr:2021vuf,Torgrimsson:2021wcj,Torgrimsson:2021zob,Ekman:2021eqc,Heinzl:2021mji},
and by LAD featuring known, non-perturbative physics: pre-accelaration and runaway solutions.
It is also a time-dependent problem, allowing us to study double expansions (in a time and a coupling), while most applications have looked at expansions in a coupling only~\cite{Florio:2019hzn,Torgrimsson:2020wlz,Mironov:2020gbi,Heinzl:2021mji,Dunne:2021acr,Borinsky:2021hnd}.
(But see Refs.~\cite{Ahmed:2017lhl,Torgrimsson:2021wcj,Torgrimsson:2021zob})

Extending our previous work on reduction of order and RR~\cite{Ekman:2021eqc} we have shown that the non-perturbative runaway solutions are eliminated by reduction of order only when an essentially perturbative initial condition is applied.
We illustrate this with the toy model (similar examples are found in several textbooks, e.g.~\cite[Ch.~7]{Bender:1999})
\begin{equation}
    \label{eq:toy-model}
    z = 1 - \varepsilon z^2
\end{equation}
for a small parameter $\varepsilon$.
The two solutions to this equation are
\begin{equation}
    \begin{split}
        z_\pm & = \frac{1}{2\varepsilon} (-1 \pm \sqrt{1 + 4 \varepsilon} ) \\
              & = \begin{cases}
                  \phantom{-\frac{1}{\varepsilon} - \;\,}
                  1 - \varepsilon + 2 \varepsilon^2 - 5 \varepsilon^3 \cdots \\
                  -\frac{1}{\varepsilon} -  1 + \varepsilon - 2 \varepsilon^2 + 5 \varepsilon^3 \cdots
              \end{cases}
              \,
              .
    \end{split}
\end{equation}
If reduction of order is initiated with $z_0 = 1 + O(\varepsilon)$ only the purely perturbative solution $z_+$ is seen.
If on the other hand an Ansatz $z_0 = c_1/\varepsilon + c_2 + O(\varepsilon)$ including a possible non-perturbative term is made,
one finds two branches $c_{i,+} = (0,1)$ and $c_{i,-} = (-1,-1)$.
These generate $z_+$ and $z_-$, respectively.
We see that it is not reduction of order itself that eliminates non-perturbative terms, but reduction of order combined with an initial condition on the purely perturbative branch.
When non-perturbative terms are large, as is the case for the toy model~\cref{eq:toy-model} and LAD, this is the only branch smoothly connected to vanishing expansion parameter.
Thus we had to set an initial condition~\cref{eq:non-pert-ic} at non-zero expansion parameter to keep non-perturbative runaway solutions with reduction of order.

We then considered the transseries structure of \emph{solutions} to LAD.
We showed to generate a solution of LAD with a given initial (or final, for pre-accelerating solutions) acceleration, instanton terms \emph{of all orders} must, in general, be kept and their initial (final) coefficients must be chosen consistently with LAD to the desired accuracy.
The one exception to this is when the initial acceleration leads to the physical, non-runaway solution: then all instanton terms vanish, and the solution is entirely perturbative.

As time-dependent quantities, solutions to LAD exemplify that expansions in two variables can display strikingly different behaviour in different regions of the variable plane and limits~\cite{Ahmed:2017lhl}, and under non-linear transformations.
First, the solution to free LAD contains non-perturbative terms of all instanton orders in one set of variables, \cemph{but in another set these are transmuted into perturbative terms}.
Secondly, in a field, both the perturbative series and the instanton series are divergent and must be resummed at small times, but convergent for large times.

\cemph{
    The coupling paramater $\delta$ is smaller than the quantum non-linearity parameter $\chi$ by a factor of $\alpha$.
    Our results for $\delta \gtrsim 1$ should therefore be read as being about classical electrodynamics as a formal theory.
    It would however be interesting to consider, e.g., quantifying how much closer $\LL\infty$, predicting less radiation reaction than $\LL1$, is to QED.
    Calculating to the necessary order in strong field QED remains extremely challenging,
    but recent progress on QED resummations and the Ritus-Narozhny conjecture~\cite{Mironov:2020gbi,Heinzl:2021mji,Torgrimsson:2021wcj,Mironov:2021bmp} offers some encouragement.
}

\cemph{Our results highlight that u}nderstanding the singularity structure of the Borel transform of a series is important for efficiently resumming it~\cite{Mera:2018qte,Costin:2019xql,Costin:2021bay}.
The series~\cref{eq:gh-sol} is difficult to resum at large, finite, times because it ``looks" convergent, with an analytic Borel transform, whereas the Pad\'e approximant has poles.
Ref.~\cite{Torgrimsson:2021zob} found that a non-linear transformation effectively performed a partial resummation in one variable leading to an expansion divergent at all times, and therefore well-suited to Borel-Pad\'e resummation.
We take this and our results as a strong indication that a more thorough understanding of multi-variable divergent expansions, Borel transforms, and transseries would be highly useful to guide resummations in time-dependent problems and other expansions in multiple parameters.

\begin{acknowledgments}
\emph{
    We thank Tom Heinzl, Anton Ilderton, and Greger Torgrimsson for useful discussions and comments on this manuscript.
    The author was supported by the Leverhulme Trust, grant RPG-2019-148.
}
\end{acknowledgments}

\appendix

\section{ \texorpdfstring{$\LL\infty$}{LL∞} in a monochromatic plane wave }
\label{app:monochromatic}

A circularly polarised monochromatic plane wave is characterised by  the wavevector $k^\mu = \omega n^\mu$,
and the invariants $\eta = k \cdot p /m^2$ and $\delta^2 = \tau_0^2 p f^2 p$.
There are only three possible tensor structures that can enter into $\LL\infty$, essentially $f, f^2, (p \cdot \partial) f$,
Because of the circular polarisation $\delta/\eta =: a_0$ is a constant and the form of $\LL\infty$ must be
\begin{equation}
    \begin{split}
        \dot{p}^\mu =
            & \CA_1(\delta) f_{\mu\nu}p^\nu + \tau_0 \CA_2(\delta) (P f^2)_{\mu\nu} p^\nu \\
            & + \tau_0 \CA_3(\delta) f_{\mu\nu,\rho} p^\nu p^\rho
            \,
            .
    \end{split}
\end{equation}

Applying reduction of order leads to the fixed-point condition analogous to~\cref{eq:fixed-point},
\begin{align}
    \label{eq:fixed-point-mc}
    \left\{\begin{array}{rcl}
        -\delta^3  \CA_2 \CA_1' - 2  \delta^2 \CA_1 \CA_2 - \frac{\delta^2}{a_0^2} \CA_3 + 1  & = & \CA_1 \\
        -\delta^3  \CA_2 \CA_2'  + \CA_1^2 - 2  \delta^2 \CA_2^2 + \frac{\delta^2}{a_0^2} \CA_3^2 & = & \CA_2 \\
        -\delta^3  \CA_2 \CA_3' - 2  \delta^2 \CA_2 \CA_3 + \CA_1 & = & \CA_3
    \end{array}
    \right.
    \,
    .
\end{align}
Note that $a_0$ enters as a parameter, but there are only derivatives with respect to $\delta$, as $a_0$ is constant.
If $a_0 \mapsto \infty, \CA_3 \mapsto 0$ the \cemph{first two equations form}~\cref{eq:fixed-point} after renaming.
\cemph{
    In this limit of a constant field, the third equation drops out: it is the coefficient in the equation of motion of the $f_{\mu\nu,\rho} p^\nu p^\rho$ term, which is then not present.
}

Starting with $\CA_i = 1 + \ordo{\delta^2}$ it is straight-forward to derive perturbative expansions in $\delta^2$.
The $\delta^{2n}$ coefficient is a degree $n$ polynomial in $1/a_0^2$;
for $\CA_{1,2}$ the $a_0^0$ term recovers the factorially divergent perturbative expansions of $\CA, \CB$ from the main text.
We therefore expect the same Borel singularity structure, and such is straightforwardly supported with the experimental, graphical methods in the main text.

To analytically determine the non-perturbative exponent we linearise the fixed-point equations~\cref{eq:fixed-point-mc} around $\CA_i = 1, 1/\delta^2 = z = \infty$, resulting in
\begin{widetext}
\begin{align}
    \frac{\ud}{\ud z}\begin{pmatrix}
        \tilde\CA_1 \\ \tilde\CA_2 \\ \tilde\CA_3
    \end{pmatrix}
    =
    \frac{1}{2}
    \underbrace{\begin{pmatrix}
         1 &  0 &  0 \\
        -2 &  1 &  0 \\
        -1 &  0 &  1 \\
    \end{pmatrix}}_{\mathbb M}
    \begin{pmatrix}
        \tilde\CA_1 \\ \tilde\CA_2 \\ \tilde\CA_3
    \end{pmatrix}
    +
    \frac{1}{2a_0^2 z}
    \underbrace{\begin{pmatrix}
        2a_0^2 &         -1 &      1 \\
             0 & 1 + 2a_0^2 &     -2 \\
             0 &          0 & 2a_0^2
    \end{pmatrix}}_{\mathbb N}
    \begin{pmatrix}
        \tilde\CA_1 \\ \tilde\CA_2 \\ \tilde\CA_3
    \end{pmatrix}
    +
    \frac{1}{2a_0^2 z} \begin{pmatrix}
        1 + 2a_0^2 \\
       -1 + 2a_0^2 \\
        2a_0^2
    \end{pmatrix}
\end{align}
The linearisation is solved by writing $\CA_i = \exp[\mathbb Mz]_{ij} \CB_j$ which leads to the equation for $\CB_i$,
\begin{equation}
    \frac{\ud \CB_i}{\ud z}
    =
    \big(e^{-\mathbb Mz} \frac{\mathbb N}{z} e^{\mathbb Mz} \big)_{ij}
    \CB_j
    +
    \frac{1}{2a_0^2} e^{-\mathbb M z}
    \begin{pmatrix}
        1 + 2a_0^2 \\
        -1 + 2a_0^2 \\
        2a_0^2
    \end{pmatrix}
    =
    \frac{1}{z}
    \begin{pmatrix}
        \CB_1 \\
        \big( 1 + \frac{1}{2a_0^2})  \CB_2 \\
        \CB_3
    \end{pmatrix}
    +
    e^{-z/2}
    \begin{pmatrix}
        1        & 0 & 0 \\
        z/a_0^2  & 1 & 0 \\
        z/2a_0^2 & 0 & 1
    \end{pmatrix}
    \begin{pmatrix}
        1 + 2a_0^2 \\
        -1 + 2a_0^2 \\
        2a_0^2
    \end{pmatrix}
    \,
    .
\end{equation}
and the general solution,
\begin{equation}
    \begin{pmatrix}
        \tilde\CA_{1} \\
        \tilde\CA_{2} \\
        \tilde\CA_{3}
    \end{pmatrix}
    =
    e^{\mathbb M z}
    \begin{pmatrix}
        c_1 z \\
        c_2 z^{1 + 1/2a_0^2} \\
        c_3 z
    \end{pmatrix}
    + \CA_{i,\text{p}}
    =
    e^{z/2}
    \begin{pmatrix}
        1 & 0 & 0 \\
        -z & 1 & 0 \\
        -z/2 & 0 & 1
    \end{pmatrix}
    \begin{pmatrix}
        c_1 z \\
        c_2 z^{1 + 1/2a_0^2} \\
        c_3 z
    \end{pmatrix}
    - \frac{1}{z}
    \begin{pmatrix}
        2 + 1/a_0^2 \\
        6 + 1/a_0^2 \\
        4 + 1/a_0^2
    \end{pmatrix}
    \,
    .
\end{equation}
\end{widetext}
We see that the same non-perturbatively \emph{large} exponential $e^{1/2\delta^2}$ appears as for the CCF.
One of the powers has an $a_0$-dependence not seen for the CCF;
this corresponds to subleading, $a_0$-dependant large-order behaviour of the perturbative coefficients.

\section{ Modified Richardson Extrapolation for Logarithmic Corrections }
\label{app:richardson}

Richardson extrapolation~\cite[Ch.~8.1]{Bender:1999} can be used to accelerate the convergence of a quantity
\begin{equation}
    f_n \sim \sum_{k \ge 0} a_k n^{-k} \xrightarrow{n \to \infty} a_0
\end{equation}
from $\ordo{1/n}$ to $\ordo{n^{-K-1}}$ for large $n$.
Specifically letting $(\Delta_n f) := f_{n+1} - f_n$ it holds that (in this Appendix all asymptotic statements are as $n \to \infty$)
\begin{equation}
    R_K[f_n] = \frac{1}{K!} (\Delta^K_n n^K f_n) = a_0 + \ordo{n^{-K-1}}
    \,
    .
\end{equation}
However as discussed by Ref.~\cite{Borinsky:2021hnd}, if the quantity $f_k$ has logarithmic corrections, viz.,
\begin{equation}
    \label{eq:log-corr}
    f_k \sim \sum_{k\ge 0} a_k n^{-k} + \log n \sum_{k \ge 1} b_k n^{-k}
\end{equation}
the acceleration is spoiled.
This is solved in Ref.~\cite{Borinsky:2021hnd} by applying $R_K$ twice such that
\begin{equation}
    R_K\big[ R_K[f_n] \big] = a_0 + \ordo{n^{-K-1} \log n}
    \,
    .
\end{equation}

There is an intuitive explanation for this.
The operator $\Delta_n$ acts like a derivative: it lowers the degree of polynomials by $1$, annihilates constants, and satisfies a quasi-Leibniz rule.
Furthermore $\Delta_n \log n = \log(1 + 1/n) = \ordo{1/n}$.
Hence in the leading term with a logarithm in~\cref{eq:log-corr}, $n^{K-1} \log n$, the ``derivative'' $\Delta_n^K$ has to act on $\log n$ at least once to produce something non-zero.
Acting another $K-1$ times produces something that goes like $n^{-1}$.
This is why logarithmic corrections spoil accelerated convergence, but $R_K[f_k]$
is itself free of logarithms.
Thus simply applying $R_K$ again kills subleading terms to order $n^{-K-1}$, as desired.

Suppose now that there are subleading terms with powers of $\log n$, viz.,
\begin{equation}
    f_k \sim \sum_{k \ge 0} a_k n^{-k} + \sum_{\ell=1}^p (\log n)^\ell \sum_{k\ge1} b_{\ell,k} n^{-k}
    \,
    .
\end{equation}
Similarly to before, $\Delta_n^K$ has to act on at least one logarithmic factor in $n^{-K-1} ( \log n)^p$ to contribute, and consequently $\Delta_n^K n^K f_k $ goes like $n^{-1} (\log n)^{p-1}$.
Iterating we realise that $R_k^p[f_n]$ is free of logarithms, and
\begin{equation}
    R_k^{p+1}[f_n] = a_0 + \ordo{ n^{-K-1} (\log n)^p }
    \,
    .
\end{equation}

\bibliography{References}

\begin{thebibliography}{65}%
\makeatletter
\providecommand \@ifxundefined [1]{%
 \@ifx{#1\undefined}
}%
\providecommand \@ifnum [1]{%
 \ifnum #1\expandafter \@firstoftwo
 \else \expandafter \@secondoftwo
 \fi
}%
\providecommand \@ifx [1]{%
 \ifx #1\expandafter \@firstoftwo
 \else \expandafter \@secondoftwo
 \fi
}%
\providecommand \natexlab [1]{#1}%
\providecommand \enquote  [1]{``#1''}%
\providecommand \bibnamefont  [1]{#1}%
\providecommand \bibfnamefont [1]{#1}%
\providecommand \citenamefont [1]{#1}%
\providecommand \href@noop [0]{\@secondoftwo}%
\providecommand \href [0]{\begingroup \@sanitize@url \@href}%
\providecommand \@href[1]{\@@startlink{#1}\@@href}%
\providecommand \@@href[1]{\endgroup#1\@@endlink}%
\providecommand \@sanitize@url [0]{\catcode `\\12\catcode `\$12\catcode
  `\&12\catcode `\#12\catcode `\^12\catcode `\_12\catcode `\%12\relax}%
\providecommand \@@startlink[1]{}%
\providecommand \@@endlink[0]{}%
\providecommand \url  [0]{\begingroup\@sanitize@url \@url }%
\providecommand \@url [1]{\endgroup\@href {#1}{\urlprefix }}%
\providecommand \urlprefix  [0]{URL }%
\providecommand \Eprint [0]{\href }%
\providecommand \doibase [0]{http://dx.doi.org/}%
\providecommand \selectlanguage [0]{\@gobble}%
\providecommand \bibinfo  [0]{\@secondoftwo}%
\providecommand \bibfield  [0]{\@secondoftwo}%
\providecommand \translation [1]{[#1]}%
\providecommand \BibitemOpen [0]{}%
\providecommand \bibitemStop [0]{}%
\providecommand \bibitemNoStop [0]{.\EOS\space}%
\providecommand \EOS [0]{\spacefactor3000\relax}%
\providecommand \BibitemShut  [1]{\csname bibitem#1\endcsname}%
\let\auto@bib@innerbib\@empty
%</preamble>
\bibitem [{\citenamefont {Cole}\ \emph {et~al.}(2018)\citenamefont {Cole},
  \citenamefont {Behm}, \citenamefont {Gerstmayr}, \citenamefont {Blackburn},
  \citenamefont {Wood}, \citenamefont {Baird}, \citenamefont {Duff},
  \citenamefont {Harvey}, \citenamefont {Ilderton}, \citenamefont {Joglekar},
  \citenamefont {Krushelnick}, \citenamefont {Kuschel}, \citenamefont
  {Marklund}, \citenamefont {McKenna}, \citenamefont {Murphy}, \citenamefont
  {Poder}, \citenamefont {Ridgers}, \citenamefont {Samarin}, \citenamefont
  {Sarri}, \citenamefont {Symes}, \citenamefont {Thomas}, \citenamefont
  {Warwick}, \citenamefont {Zepf}, \citenamefont {Najmudin},\ and\
  \citenamefont {Mangles}}]{Cole:2017zca}%
  \BibitemOpen
  \bibfield  {author} {\bibinfo {author} {\bibfnamefont {J.~M.}\ \bibnamefont
  {Cole}}, \bibinfo {author} {\bibfnamefont {K.~T.}\ \bibnamefont {Behm}},
  \bibinfo {author} {\bibfnamefont {E.}~\bibnamefont {Gerstmayr}}, \bibinfo
  {author} {\bibfnamefont {T.~G.}\ \bibnamefont {Blackburn}}, \bibinfo {author}
  {\bibfnamefont {J.~C.}\ \bibnamefont {Wood}}, \bibinfo {author}
  {\bibfnamefont {C.~D.}\ \bibnamefont {Baird}}, \bibinfo {author}
  {\bibfnamefont {M.~J.}\ \bibnamefont {Duff}}, \bibinfo {author}
  {\bibfnamefont {C.}~\bibnamefont {Harvey}}, \bibinfo {author} {\bibfnamefont
  {A.}~\bibnamefont {Ilderton}}, \bibinfo {author} {\bibfnamefont {A.~S.}\
  \bibnamefont {Joglekar}}, \bibinfo {author} {\bibfnamefont {K.}~\bibnamefont
  {Krushelnick}}, \bibinfo {author} {\bibfnamefont {S.}~\bibnamefont
  {Kuschel}}, \bibinfo {author} {\bibfnamefont {M.}~\bibnamefont {Marklund}},
  \bibinfo {author} {\bibfnamefont {P.}~\bibnamefont {McKenna}}, \bibinfo
  {author} {\bibfnamefont {C.~D.}\ \bibnamefont {Murphy}}, \bibinfo {author}
  {\bibfnamefont {K.}~\bibnamefont {Poder}}, \bibinfo {author} {\bibfnamefont
  {C.~P.}\ \bibnamefont {Ridgers}}, \bibinfo {author} {\bibfnamefont {G.~M.}\
  \bibnamefont {Samarin}}, \bibinfo {author} {\bibfnamefont {G.}~\bibnamefont
  {Sarri}}, \bibinfo {author} {\bibfnamefont {D.~R.}\ \bibnamefont {Symes}},
  \bibinfo {author} {\bibfnamefont {A.~G.~R.}\ \bibnamefont {Thomas}}, \bibinfo
  {author} {\bibfnamefont {J.}~\bibnamefont {Warwick}}, \bibinfo {author}
  {\bibfnamefont {M.}~\bibnamefont {Zepf}}, \bibinfo {author} {\bibfnamefont
  {Z.}~\bibnamefont {Najmudin}}, \ and\ \bibinfo {author} {\bibfnamefont
  {S.~P.~D.}\ \bibnamefont {Mangles}},\ }\href {\doibase
  10.1103/PhysRevX.8.011020} {\bibfield  {journal} {\bibinfo  {journal} {Phys.
  Rev. X}\ }\textbf {\bibinfo {volume} {8}},\ \bibinfo {pages} {011020}
  (\bibinfo {year} {2018})},\ \Eprint {http://arxiv.org/abs/1707.06821}
  {arXiv:1707.06821 [physics.plasm-ph]} \BibitemShut {NoStop}%
\bibitem [{\citenamefont {{Poder}}\ \emph {et~al.}(2018)\citenamefont
  {{Poder}}, \citenamefont {{Tamburini}}, \citenamefont {{Sarri}},
  \citenamefont {{Di Piazza}}, \citenamefont {{Kuschel}}, \citenamefont
  {{Baird}}, \citenamefont {{Behm}}, \citenamefont {{Bohlen}}, \citenamefont
  {{Cole}}, \citenamefont {{Corvan}}, \citenamefont {{Duff}}, \citenamefont
  {{Gerstmayr}}, \citenamefont {{Keitel}}, \citenamefont {{Krushelnick}},
  \citenamefont {{Mangles}}, \citenamefont {{McKenna}}, \citenamefont
  {{Murphy}}, \citenamefont {{Najmudin}}, \citenamefont {{Ridgers}},
  \citenamefont {{Samarin}}, \citenamefont {{Symes}}, \citenamefont {{Thomas}},
  \citenamefont {{Warwick}},\ and\ \citenamefont {{Zepf}}}]{Poder:2018ifi}%
  \BibitemOpen
  \bibfield  {author} {\bibinfo {author} {\bibfnamefont {K.}~\bibnamefont
  {{Poder}}}, \bibinfo {author} {\bibfnamefont {M.}~\bibnamefont
  {{Tamburini}}}, \bibinfo {author} {\bibfnamefont {G.}~\bibnamefont
  {{Sarri}}}, \bibinfo {author} {\bibfnamefont {A.}~\bibnamefont {{Di
  Piazza}}}, \bibinfo {author} {\bibfnamefont {S.}~\bibnamefont {{Kuschel}}},
  \bibinfo {author} {\bibfnamefont {C.~D.}\ \bibnamefont {{Baird}}}, \bibinfo
  {author} {\bibfnamefont {K.}~\bibnamefont {{Behm}}}, \bibinfo {author}
  {\bibfnamefont {S.}~\bibnamefont {{Bohlen}}}, \bibinfo {author}
  {\bibfnamefont {J.~M.}\ \bibnamefont {{Cole}}}, \bibinfo {author}
  {\bibfnamefont {D.~J.}\ \bibnamefont {{Corvan}}}, \bibinfo {author}
  {\bibfnamefont {M.}~\bibnamefont {{Duff}}}, \bibinfo {author} {\bibfnamefont
  {E.}~\bibnamefont {{Gerstmayr}}}, \bibinfo {author} {\bibfnamefont {C.~H.}\
  \bibnamefont {{Keitel}}}, \bibinfo {author} {\bibfnamefont {K.}~\bibnamefont
  {{Krushelnick}}}, \bibinfo {author} {\bibfnamefont {S.~P.~D.}\ \bibnamefont
  {{Mangles}}}, \bibinfo {author} {\bibfnamefont {P.}~\bibnamefont
  {{McKenna}}}, \bibinfo {author} {\bibfnamefont {C.~D.}\ \bibnamefont
  {{Murphy}}}, \bibinfo {author} {\bibfnamefont {Z.}~\bibnamefont
  {{Najmudin}}}, \bibinfo {author} {\bibfnamefont {C.~P.}\ \bibnamefont
  {{Ridgers}}}, \bibinfo {author} {\bibfnamefont {G.~M.}\ \bibnamefont
  {{Samarin}}}, \bibinfo {author} {\bibfnamefont {D.~R.}\ \bibnamefont
  {{Symes}}}, \bibinfo {author} {\bibfnamefont {A.~G.~R.}\ \bibnamefont
  {{Thomas}}}, \bibinfo {author} {\bibfnamefont {J.}~\bibnamefont {{Warwick}}},
  \ and\ \bibinfo {author} {\bibfnamefont {M.}~\bibnamefont {{Zepf}}},\ }\href
  {\doibase 10.1103/PhysRevX.8.031004} {\bibfield  {journal} {\bibinfo
  {journal} {Phys. Rev. X}\ }\textbf {\bibinfo {volume} {8}},\ \bibinfo {pages}
  {031004} (\bibinfo {year} {2018})},\ \Eprint
  {http://arxiv.org/abs/1709.01861} {arXiv:1709.01861 [physics.plasm-ph]}
  \BibitemShut {NoStop}%
\bibitem [{\citenamefont {Burton}\ and\ \citenamefont
  {Noble}(2014)}]{Burton:2014wsa}%
  \BibitemOpen
  \bibfield  {author} {\bibinfo {author} {\bibfnamefont {D.~A.}\ \bibnamefont
  {Burton}}\ and\ \bibinfo {author} {\bibfnamefont {A.}~\bibnamefont {Noble}},\
  }\href {\doibase 10.1080/00107514.2014.886840} {\bibfield  {journal}
  {\bibinfo  {journal} {Contemp. Phys.}\ }\textbf {\bibinfo {volume} {55}},\
  \bibinfo {pages} {110} (\bibinfo {year} {2014})},\ \Eprint
  {http://arxiv.org/abs/1409.7707} {arXiv:1409.7707 [physics.plasm-ph]}
  \BibitemShut {NoStop}%
\bibitem [{\citenamefont {Blackburn}(2020)}]{Blackburn:2019rfv}%
  \BibitemOpen
  \bibfield  {author} {\bibinfo {author} {\bibfnamefont {T.~G.}\ \bibnamefont
  {Blackburn}},\ }\href {\doibase 10.1007/s41614-020-0042-0} {\bibfield
  {journal} {\bibinfo  {journal} {Rev. Mod. Plasma Phys.}\ }\textbf {\bibinfo
  {volume} {4}},\ \bibinfo {pages} {5} (\bibinfo {year} {2020})},\ \Eprint
  {http://arxiv.org/abs/1910.13377} {arXiv:1910.13377 [physics.plasm-ph]}
  \BibitemShut {NoStop}%
\bibitem [{\citenamefont {Gonoskov}\ \emph {et~al.}(2021)\citenamefont
  {Gonoskov}, \citenamefont {Blackburn}, \citenamefont {Marklund},\ and\
  \citenamefont {Bulanov}}]{Gonoskov:2021hwf}%
  \BibitemOpen
  \bibfield  {author} {\bibinfo {author} {\bibfnamefont {A.}~\bibnamefont
  {Gonoskov}}, \bibinfo {author} {\bibfnamefont {T.~G.}\ \bibnamefont
  {Blackburn}}, \bibinfo {author} {\bibfnamefont {M.}~\bibnamefont {Marklund}},
  \ and\ \bibinfo {author} {\bibfnamefont {S.~S.}\ \bibnamefont {Bulanov}},\
  }\href@noop {} {\  (\bibinfo {year} {2021})},\ \Eprint
  {http://arxiv.org/abs/2107.02161} {arXiv:2107.02161 [physics.plasm-ph]}
  \BibitemShut {NoStop}%
\bibitem [{\citenamefont {Danson}\ \emph {et~al.}(2019)\citenamefont {Danson}
  \emph {et~al.}}]{danson2019petawatt}%
  \BibitemOpen
  \bibfield  {author} {\bibinfo {author} {\bibfnamefont {C.~N.}\ \bibnamefont
  {Danson}} \emph {et~al.},\ }\href {\doibase 10.1017/hpl.2014.52} {\bibfield
  {journal} {\bibinfo  {journal} {High Power Laser Sci. Eng.}\ }\textbf
  {\bibinfo {volume} {7}},\ \bibinfo {pages} {e54} (\bibinfo {year}
  {2019})}\BibitemShut {NoStop}%
\bibitem [{\citenamefont {Abramowicz}\ \emph {et~al.}(2021)\citenamefont
  {Abramowicz} \emph {et~al.}}]{Abramowicz:2021zja}%
  \BibitemOpen
  \bibfield  {author} {\bibinfo {author} {\bibfnamefont {H.}~\bibnamefont
  {Abramowicz}} \emph {et~al.},\ }\href@noop {} {\enquote {\bibinfo {title}
  {{Conceptual Design Report for the LUXE Experiment}},}\ } (\bibinfo {year}
  {2021}),\ \Eprint {http://arxiv.org/abs/2102.02032} {arXiv:2102.02032
  [hep-ex]} \BibitemShut {NoStop}%
\bibitem [{\citenamefont {Meuren}\ \emph {et~al.}(2020)\citenamefont {Meuren}
  \emph {et~al.}}]{Meuren:2020nbw}%
  \BibitemOpen
  \bibfield  {author} {\bibinfo {author} {\bibfnamefont {S.}~\bibnamefont
  {Meuren}} \emph {et~al.},\ }\href@noop {} {\enquote {\bibinfo {title} {{On
  Seminal HEDP Research Opportunities Enabled by Colocating Multi-Petawatt
  Laser with High-Density Electron Beams}},}\ } (\bibinfo {year} {2020}),\
  \Eprint {http://arxiv.org/abs/2002.10051} {arXiv:2002.10051
  [physics.plasm-ph]} \BibitemShut {NoStop}%
\bibitem [{\citenamefont {Taylor}\ \emph {et~al.}(1979)\citenamefont {Taylor},
  \citenamefont {Fowler},\ and\ \citenamefont {McCulloch}}]{Taylor:1979zz}%
  \BibitemOpen
  \bibfield  {author} {\bibinfo {author} {\bibfnamefont {J.~H.}\ \bibnamefont
  {Taylor}}, \bibinfo {author} {\bibfnamefont {L.~A.}\ \bibnamefont {Fowler}},
  \ and\ \bibinfo {author} {\bibfnamefont {P.~M.}\ \bibnamefont {McCulloch}},\
  }\href {\doibase 10.1038/277437a0} {\bibfield  {journal} {\bibinfo  {journal}
  {Nature}\ }\textbf {\bibinfo {volume} {277}},\ \bibinfo {pages} {437}
  (\bibinfo {year} {1979})}\BibitemShut {NoStop}%
\bibitem [{\citenamefont {Damour}(2020)}]{Damour:2020tta}%
  \BibitemOpen
  \bibfield  {author} {\bibinfo {author} {\bibfnamefont {T.}~\bibnamefont
  {Damour}},\ }\href {\doibase 10.1103/PhysRevD.102.124008} {\bibfield
  {journal} {\bibinfo  {journal} {Phys. Rev. D}\ }\textbf {\bibinfo {volume}
  {102}},\ \bibinfo {pages} {124008} (\bibinfo {year} {2020})},\ \Eprint
  {http://arxiv.org/abs/2010.01641} {arXiv:2010.01641 [gr-qc]} \BibitemShut
  {NoStop}%
\bibitem [{\citenamefont {Di~Vecchia}\ \emph {et~al.}(2021)\citenamefont
  {Di~Vecchia}, \citenamefont {Heissenberg}, \citenamefont {Russo},\ and\
  \citenamefont {Veneziano}}]{DiVecchia:2021bdo}%
  \BibitemOpen
  \bibfield  {author} {\bibinfo {author} {\bibfnamefont {P.}~\bibnamefont
  {Di~Vecchia}}, \bibinfo {author} {\bibfnamefont {C.}~\bibnamefont
  {Heissenberg}}, \bibinfo {author} {\bibfnamefont {R.}~\bibnamefont {Russo}},
  \ and\ \bibinfo {author} {\bibfnamefont {G.}~\bibnamefont {Veneziano}},\
  }\href@noop {} {\  (\bibinfo {year} {2021})},\ \Eprint
  {http://arxiv.org/abs/2104.03256} {arXiv:2104.03256 [hep-th]} \BibitemShut
  {NoStop}%
\bibitem [{\citenamefont {Herrmann}\ \emph {et~al.}(2021)\citenamefont
  {Herrmann}, \citenamefont {Parra-Martinez}, \citenamefont {Ruf},\ and\
  \citenamefont {Zeng}}]{Herrmann:2021tct}%
  \BibitemOpen
  \bibfield  {author} {\bibinfo {author} {\bibfnamefont {E.}~\bibnamefont
  {Herrmann}}, \bibinfo {author} {\bibfnamefont {J.}~\bibnamefont
  {Parra-Martinez}}, \bibinfo {author} {\bibfnamefont {M.~S.}\ \bibnamefont
  {Ruf}}, \ and\ \bibinfo {author} {\bibfnamefont {M.}~\bibnamefont {Zeng}},\
  }\href@noop {} {\  (\bibinfo {year} {2021})},\ \Eprint
  {http://arxiv.org/abs/2104.03957} {arXiv:2104.03957 [hep-th]} \BibitemShut
  {NoStop}%
\bibitem [{\citenamefont {Bjerrum-Bohr}\ \emph {et~al.}(2021)\citenamefont
  {Bjerrum-Bohr}, \citenamefont {Damgaard}, \citenamefont {Plant\'e},\ and\
  \citenamefont {Vanhove}}]{Bjerrum-Bohr:2021vuf}%
  \BibitemOpen
  \bibfield  {author} {\bibinfo {author} {\bibfnamefont {N.~E.~J.}\
  \bibnamefont {Bjerrum-Bohr}}, \bibinfo {author} {\bibfnamefont {P.~H.}\
  \bibnamefont {Damgaard}}, \bibinfo {author} {\bibfnamefont {L.}~\bibnamefont
  {Plant\'e}}, \ and\ \bibinfo {author} {\bibfnamefont {P.}~\bibnamefont
  {Vanhove}},\ }\href@noop {} {\  (\bibinfo {year} {2021})},\ \Eprint
  {http://arxiv.org/abs/2104.04510} {arXiv:2104.04510 [hep-th]} \BibitemShut
  {NoStop}%
\bibitem [{\citenamefont
  {Torgrimsson}(2020{\natexlab{a}})}]{Torgrimsson:2020mto}%
  \BibitemOpen
  \bibfield  {author} {\bibinfo {author} {\bibfnamefont {G.}~\bibnamefont
  {Torgrimsson}},\ }\href {\doibase 10.1103/PhysRevD.102.116008} {\bibfield
  {journal} {\bibinfo  {journal} {Phys. Rev. D}\ }\textbf {\bibinfo {volume}
  {102}},\ \bibinfo {pages} {116008} (\bibinfo {year} {2020}{\natexlab{a}})},\
  \Eprint {http://arxiv.org/abs/2010.02128} {arXiv:2010.02128 [hep-ph]}
  \BibitemShut {NoStop}%
\bibitem [{\citenamefont
  {Torgrimsson}(2020{\natexlab{b}})}]{Torgrimsson:2020wlz}%
  \BibitemOpen
  \bibfield  {author} {\bibinfo {author} {\bibfnamefont {G.}~\bibnamefont
  {Torgrimsson}},\ }\href {\doibase 10.1103/PhysRevD.102.096008} {\bibfield
  {journal} {\bibinfo  {journal} {Phys. Rev. D}\ }\textbf {\bibinfo {volume}
  {102}},\ \bibinfo {pages} {096008} (\bibinfo {year} {2020}{\natexlab{b}})},\
  \Eprint {http://arxiv.org/abs/2007.08492} {arXiv:2007.08492 [hep-ph]}
  \BibitemShut {NoStop}%
\bibitem [{\citenamefont {Karbstein}(2019)}]{Karbstein:2019wmj}%
  \BibitemOpen
  \bibfield  {author} {\bibinfo {author} {\bibfnamefont {F.}~\bibnamefont
  {Karbstein}},\ }\href {\doibase 10.1103/PhysRevLett.122.211602} {\bibfield
  {journal} {\bibinfo  {journal} {Phys. Rev. Lett.}\ }\textbf {\bibinfo
  {volume} {122}},\ \bibinfo {pages} {211602} (\bibinfo {year} {2019})},\
  \Eprint {http://arxiv.org/abs/1903.06998} {arXiv:1903.06998 [hep-th]}
  \BibitemShut {NoStop}%
\bibitem [{\citenamefont {Mironov}\ \emph {et~al.}(2020)\citenamefont
  {Mironov}, \citenamefont {Meuren},\ and\ \citenamefont
  {Fedotov}}]{Mironov:2020gbi}%
  \BibitemOpen
  \bibfield  {author} {\bibinfo {author} {\bibfnamefont {A.~A.}\ \bibnamefont
  {Mironov}}, \bibinfo {author} {\bibfnamefont {S.}~\bibnamefont {Meuren}}, \
  and\ \bibinfo {author} {\bibfnamefont {A.~M.}\ \bibnamefont {Fedotov}},\
  }\href {\doibase 10.1103/PhysRevD.102.053005} {\bibfield  {journal} {\bibinfo
   {journal} {Phys. Rev. D}\ }\textbf {\bibinfo {volume} {102}},\ \bibinfo
  {pages} {053005} (\bibinfo {year} {2020})},\ \Eprint
  {http://arxiv.org/abs/2003.06909} {arXiv:2003.06909 [hep-th]} \BibitemShut
  {NoStop}%
\bibitem [{\citenamefont {Heinzl}\ \emph {et~al.}(2021)\citenamefont {Heinzl},
  \citenamefont {Ilderton},\ and\ \citenamefont {King}}]{Heinzl:2021mji}%
  \BibitemOpen
  \bibfield  {author} {\bibinfo {author} {\bibfnamefont {T.}~\bibnamefont
  {Heinzl}}, \bibinfo {author} {\bibfnamefont {A.}~\bibnamefont {Ilderton}}, \
  and\ \bibinfo {author} {\bibfnamefont {B.}~\bibnamefont {King}},\ }\href
  {\doibase 10.1103/PhysRevLett.127.061601} {\bibfield  {journal} {\bibinfo
  {journal} {Phys. Rev. Lett.}\ }\textbf {\bibinfo {volume} {127}},\ \bibinfo
  {pages} {061601} (\bibinfo {year} {2021})},\ \Eprint
  {http://arxiv.org/abs/2101.12111} {arXiv:2101.12111 [hep-ph]} \BibitemShut
  {NoStop}%
\bibitem [{\citenamefont
  {Torgrimsson}(2021{\natexlab{a}})}]{Torgrimsson:2021wcj}%
  \BibitemOpen
  \bibfield  {author} {\bibinfo {author} {\bibfnamefont {G.}~\bibnamefont
  {Torgrimsson}},\ }\href {\doibase 10.1103/PhysRevLett.127.111602} {\bibfield
  {journal} {\bibinfo  {journal} {Phys. Rev. Lett.}\ }\textbf {\bibinfo
  {volume} {127}},\ \bibinfo {pages} {111602} (\bibinfo {year}
  {2021}{\natexlab{a}})},\ \Eprint {http://arxiv.org/abs/2102.11346}
  {arXiv:2102.11346 [hep-ph]} \BibitemShut {NoStop}%
\bibitem [{\citenamefont {Ekman}\ \emph
  {et~al.}(2021{\natexlab{a}})\citenamefont {Ekman}, \citenamefont {Heinzl},\
  and\ \citenamefont {Ilderton}}]{Ekman:2021eqc}%
  \BibitemOpen
  \bibfield  {author} {\bibinfo {author} {\bibfnamefont {R.}~\bibnamefont
  {Ekman}}, \bibinfo {author} {\bibfnamefont {T.}~\bibnamefont {Heinzl}}, \
  and\ \bibinfo {author} {\bibfnamefont {A.}~\bibnamefont {Ilderton}},\ }\href
  {\doibase 10.1103/PhysRevD.104.036002} {\bibfield  {journal} {\bibinfo
  {journal} {Phys. Rev. D}\ }\textbf {\bibinfo {volume} {104}},\ \bibinfo
  {pages} {036002} (\bibinfo {year} {2021}{\natexlab{a}})},\ \Eprint
  {http://arxiv.org/abs/2105.01640} {arXiv:2105.01640 [hep-ph]} \BibitemShut
  {NoStop}%
\bibitem [{\citenamefont
  {Torgrimsson}(2021{\natexlab{b}})}]{Torgrimsson:2021zob}%
  \BibitemOpen
  \bibfield  {author} {\bibinfo {author} {\bibfnamefont {G.}~\bibnamefont
  {Torgrimsson}},\ }\href {\doibase 10.1103/PhysRevD.104.056016} {\bibfield
  {journal} {\bibinfo  {journal} {Phys. Rev. D}\ }\textbf {\bibinfo {volume}
  {104}},\ \bibinfo {pages} {056016} (\bibinfo {year} {2021}{\natexlab{b}})},\
  \Eprint {http://arxiv.org/abs/2105.02220} {arXiv:2105.02220 [hep-ph]}
  \BibitemShut {NoStop}%
\bibitem [{\citenamefont {Taya}\ \emph
  {et~al.}(2021{\natexlab{a}})\citenamefont {Taya}, \citenamefont {Fujimori},
  \citenamefont {Misumi}, \citenamefont {Nitta},\ and\ \citenamefont
  {Sakai}}]{Taya:2020dco}%
  \BibitemOpen
  \bibfield  {author} {\bibinfo {author} {\bibfnamefont {H.}~\bibnamefont
  {Taya}}, \bibinfo {author} {\bibfnamefont {T.}~\bibnamefont {Fujimori}},
  \bibinfo {author} {\bibfnamefont {T.}~\bibnamefont {Misumi}}, \bibinfo
  {author} {\bibfnamefont {M.}~\bibnamefont {Nitta}}, \ and\ \bibinfo {author}
  {\bibfnamefont {N.}~\bibnamefont {Sakai}},\ }\href {\doibase
  10.1007/JHEP03(2021)082} {\bibfield  {journal} {\bibinfo  {journal} {JHEP}\
  }\textbf {\bibinfo {volume} {03}},\ \bibinfo {pages} {082} (\bibinfo {year}
  {2021}{\natexlab{a}})},\ \Eprint {http://arxiv.org/abs/2010.16080}
  {arXiv:2010.16080 [hep-th]} \BibitemShut {NoStop}%
\bibitem [{\citenamefont {Taya}\ \emph
  {et~al.}(2021{\natexlab{b}})\citenamefont {Taya}, \citenamefont {Hongo},\
  and\ \citenamefont {Ikeda}}]{Taya:2021dcz}%
  \BibitemOpen
  \bibfield  {author} {\bibinfo {author} {\bibfnamefont {H.}~\bibnamefont
  {Taya}}, \bibinfo {author} {\bibfnamefont {M.}~\bibnamefont {Hongo}}, \ and\
  \bibinfo {author} {\bibfnamefont {T.~N.}\ \bibnamefont {Ikeda}},\ }\href
  {\doibase 10.1103/PhysRevB.104.L140305} {\bibfield  {journal} {\bibinfo
  {journal} {Phys. Rev. B}\ }\textbf {\bibinfo {volume} {104}},\ \bibinfo
  {pages} {L140305} (\bibinfo {year} {2021}{\natexlab{b}})},\ \Eprint
  {http://arxiv.org/abs/2105.12446} {arXiv:2105.12446 [cond-mat.str-el]}
  \BibitemShut {NoStop}%
\bibitem [{\citenamefont {Dunne}\ and\ \citenamefont
  {Harris}(2021)}]{Dunne:2021acr}%
  \BibitemOpen
  \bibfield  {author} {\bibinfo {author} {\bibfnamefont {G.~V.}\ \bibnamefont
  {Dunne}}\ and\ \bibinfo {author} {\bibfnamefont {Z.}~\bibnamefont {Harris}},\
  }\href {\doibase 10.1103/PhysRevD.103.065015} {\bibfield  {journal} {\bibinfo
   {journal} {Phys. Rev. D}\ }\textbf {\bibinfo {volume} {103}},\ \bibinfo
  {pages} {065015} (\bibinfo {year} {2021})},\ \Eprint
  {http://arxiv.org/abs/2101.10409} {arXiv:2101.10409 [hep-th]} \BibitemShut
  {NoStop}%
\bibitem [{\citenamefont {Bender}\ and\ \citenamefont
  {Orszag}(1999)}]{Bender:1999}%
  \BibitemOpen
  \bibfield  {author} {\bibinfo {author} {\bibfnamefont {C.}~\bibnamefont
  {Bender}}\ and\ \bibinfo {author} {\bibfnamefont {S.}~\bibnamefont
  {Orszag}},\ }\href@noop {} {\emph {\bibinfo {title} {Advanced Mathematical
  Methods for Scientists and Engineers}}}\ (\bibinfo  {publisher} {Springer,
  New York},\ \bibinfo {year} {1999})\BibitemShut {NoStop}%
\bibitem [{\citenamefont {Mari\~no}(2014)}]{Marino:2012zq}%
  \BibitemOpen
  \bibfield  {author} {\bibinfo {author} {\bibfnamefont {M.}~\bibnamefont
  {Mari\~no}},\ }\href {\doibase 10.1002/prop.201400005} {\bibfield  {journal}
  {\bibinfo  {journal} {Fortsch. Phys.}\ }\textbf {\bibinfo {volume} {62}},\
  \bibinfo {pages} {455} (\bibinfo {year} {2014})},\ \Eprint
  {http://arxiv.org/abs/1206.6272} {arXiv:1206.6272 [hep-th]} \BibitemShut
  {NoStop}%
\bibitem [{\citenamefont {Dorigoni}(2019)}]{Dorigoni:2014hea}%
  \BibitemOpen
  \bibfield  {author} {\bibinfo {author} {\bibfnamefont {D.}~\bibnamefont
  {Dorigoni}},\ }\href {\doibase 10.1016/j.aop.2019.167914} {\bibfield
  {journal} {\bibinfo  {journal} {Annals Phys.}\ }\textbf {\bibinfo {volume}
  {409}},\ \bibinfo {pages} {167914} (\bibinfo {year} {2019})},\ \Eprint
  {http://arxiv.org/abs/1411.3585} {arXiv:1411.3585 [hep-th]} \BibitemShut
  {NoStop}%
\bibitem [{\citenamefont {Dunne}\ and\ \citenamefont
  {\"Unsal}(2016)}]{Dunne:2015eaa}%
  \BibitemOpen
  \bibfield  {author} {\bibinfo {author} {\bibfnamefont {G.~V.}\ \bibnamefont
  {Dunne}}\ and\ \bibinfo {author} {\bibfnamefont {M.}~\bibnamefont
  {\"Unsal}},\ }\href {\doibase 10.22323/1.251.0010} {\bibfield  {journal}
  {\bibinfo  {journal} {PoS}\ }\textbf {\bibinfo {volume} {LATTICE2015}},\
  \bibinfo {pages} {010} (\bibinfo {year} {2016})},\ \Eprint
  {http://arxiv.org/abs/1511.05977} {arXiv:1511.05977 [hep-lat]} \BibitemShut
  {NoStop}%
\bibitem [{\citenamefont {Aniceto}\ \emph {et~al.}(2019)\citenamefont
  {Aniceto}, \citenamefont {Basar},\ and\ \citenamefont
  {Schiappa}}]{Aniceto:2018bis}%
  \BibitemOpen
  \bibfield  {author} {\bibinfo {author} {\bibfnamefont {I.}~\bibnamefont
  {Aniceto}}, \bibinfo {author} {\bibfnamefont {G.}~\bibnamefont {Basar}}, \
  and\ \bibinfo {author} {\bibfnamefont {R.}~\bibnamefont {Schiappa}},\ }\href
  {\doibase 10.1016/j.physrep.2019.02.003} {\bibfield  {journal} {\bibinfo
  {journal} {Phys. Rept.}\ }\textbf {\bibinfo {volume} {809}},\ \bibinfo
  {pages} {1} (\bibinfo {year} {2019})},\ \Eprint
  {http://arxiv.org/abs/1802.10441} {arXiv:1802.10441 [hep-th]} \BibitemShut
  {NoStop}%
\bibitem [{\citenamefont {Costin}\ and\ \citenamefont
  {Dunne}(2019)}]{Costin:2019xql}%
  \BibitemOpen
  \bibfield  {author} {\bibinfo {author} {\bibfnamefont {O.}~\bibnamefont
  {Costin}}\ and\ \bibinfo {author} {\bibfnamefont {G.~V.}\ \bibnamefont
  {Dunne}},\ }\href {\doibase 10.1088/1751-8121/ab477b} {\bibfield  {journal}
  {\bibinfo  {journal} {J. Phys. A}\ }\textbf {\bibinfo {volume} {52}},\
  \bibinfo {pages} {445205} (\bibinfo {year} {2019})},\ \Eprint
  {http://arxiv.org/abs/1904.11593} {arXiv:1904.11593 [hep-th]} \BibitemShut
  {NoStop}%
\bibitem [{\citenamefont {Costin}\ and\ \citenamefont
  {Dunne}(2020)}]{Costin:2020hwg}%
  \BibitemOpen
  \bibfield  {author} {\bibinfo {author} {\bibfnamefont {O.}~\bibnamefont
  {Costin}}\ and\ \bibinfo {author} {\bibfnamefont {G.~V.}\ \bibnamefont
  {Dunne}},\ }\href {\doibase 10.1016/j.physletb.2020.135627} {\bibfield
  {journal} {\bibinfo  {journal} {Phys. Lett. B}\ }\textbf {\bibinfo {volume}
  {808}},\ \bibinfo {pages} {135627} (\bibinfo {year} {2020})},\ \Eprint
  {http://arxiv.org/abs/2003.07451} {arXiv:2003.07451 [hep-th]} \BibitemShut
  {NoStop}%
\bibitem [{\citenamefont {Narozhny}(1980)}]{PhysRevD.21.1176}%
  \BibitemOpen
  \bibfield  {author} {\bibinfo {author} {\bibfnamefont {N.~B.}\ \bibnamefont
  {Narozhny}},\ }\href {\doibase 10.1103/PhysRevD.21.1176} {\bibfield
  {journal} {\bibinfo  {journal} {Phys. Rev. D}\ }\textbf {\bibinfo {volume}
  {21}},\ \bibinfo {pages} {1176} (\bibinfo {year} {1980})}\BibitemShut
  {NoStop}%
\bibitem [{\citenamefont {Fedotov}(2017)}]{Fedotov:2016afw}%
  \BibitemOpen
  \bibfield  {author} {\bibinfo {author} {\bibfnamefont {A.~M.}\ \bibnamefont
  {Fedotov}},\ }\href {\doibase 10.1088/1742-6596/826/1/012027} {\bibfield
  {journal} {\bibinfo  {journal} {J. Phys. Conf. Ser.}\ }\textbf {\bibinfo
  {volume} {826}},\ \bibinfo {pages} {012027} (\bibinfo {year} {2017})},\
  \Eprint {http://arxiv.org/abs/1608.02261} {arXiv:1608.02261 [hep-ph]}
  \BibitemShut {NoStop}%
\bibitem [{\citenamefont {Abraham}(1905)}]{abraham1905}%
  \BibitemOpen
  \bibfield  {author} {\bibinfo {author} {\bibfnamefont {M.}~\bibnamefont
  {Abraham}},\ }\href@noop {} {\emph {\bibinfo {title} {Theorie der
  {E}lektrizit{\"a}t}}}\ (\bibinfo  {publisher} {Teubner},\ \bibinfo {address}
  {Leipzig},\ \bibinfo {year} {1905})\BibitemShut {NoStop}%
\bibitem [{\citenamefont {Lorentz}(1909)}]{lorentz1909}%
  \BibitemOpen
  \bibfield  {author} {\bibinfo {author} {\bibfnamefont {H.~A.}\ \bibnamefont
  {Lorentz}},\ }\href@noop {} {\emph {\bibinfo {title} {The Theory of
  Electrons}}}\ (\bibinfo  {publisher} {Teubner},\ \bibinfo {address}
  {Leipzig},\ \bibinfo {year} {1909})\BibitemShut {NoStop}%
\bibitem [{\citenamefont {Dirac}(1938)}]{dirac1938classical}%
  \BibitemOpen
  \bibfield  {author} {\bibinfo {author} {\bibfnamefont {P.~A.~M.}\
  \bibnamefont {Dirac}},\ }\href {\doibase 10.1098/rspa.193} {\bibfield
  {journal} {\bibinfo  {journal} {Proc. Roy. Soc. A}\ }\textbf {\bibinfo
  {volume} {167}},\ \bibinfo {pages} {148} (\bibinfo {year}
  {1938})}\BibitemShut {NoStop}%
\bibitem [{\citenamefont {Landau}\ and\ \citenamefont
  {Lifshitz}(1975)}]{LandauLifshitzII}%
  \BibitemOpen
  \bibfield  {author} {\bibinfo {author} {\bibfnamefont {L.~D.}\ \bibnamefont
  {Landau}}\ and\ \bibinfo {author} {\bibfnamefont {E.~M.}\ \bibnamefont
  {Lifshitz}},\ }\href@noop {} {\emph {\bibinfo {title} {The Classical Theory
  of Fields}}}\ (\bibinfo  {publisher} {Butterworth-Heinemann},\ \bibinfo
  {year} {1975})\BibitemShut {NoStop}%
\bibitem [{\citenamefont {Di~Piazza}\ \emph {et~al.}(2012)\citenamefont
  {Di~Piazza}, \citenamefont {Muller}, \citenamefont {Hatsagortsyan},\ and\
  \citenamefont {Keitel}}]{DiPiazza:2011tq}%
  \BibitemOpen
  \bibfield  {author} {\bibinfo {author} {\bibfnamefont {A.}~\bibnamefont
  {Di~Piazza}}, \bibinfo {author} {\bibfnamefont {C.}~\bibnamefont {Muller}},
  \bibinfo {author} {\bibfnamefont {K.~Z.}\ \bibnamefont {Hatsagortsyan}}, \
  and\ \bibinfo {author} {\bibfnamefont {C.~H.}\ \bibnamefont {Keitel}},\
  }\href {\doibase 10.1103/RevModPhys.84.1177} {\bibfield  {journal} {\bibinfo
  {journal} {Rev. Mod. Phys.}\ }\textbf {\bibinfo {volume} {84}},\ \bibinfo
  {pages} {1177} (\bibinfo {year} {2012})},\ \Eprint
  {http://arxiv.org/abs/1111.3886} {arXiv:1111.3886 [hep-ph]} \BibitemShut
  {NoStop}%
\bibitem [{\citenamefont {Di~Piazza}(2008)}]{PiazzaExact}%
  \BibitemOpen
  \bibfield  {author} {\bibinfo {author} {\bibfnamefont {A.}~\bibnamefont
  {Di~Piazza}},\ }\href {\doibase 10.1007/s11005-008-0228-9} {\bibfield
  {journal} {\bibinfo  {journal} {{Lett. Math. Phys.}}\ }\textbf {\bibinfo
  {volume} {83}},\ \bibinfo {pages} {305} (\bibinfo {year} {2008})}\BibitemShut
  {NoStop}%
\bibitem [{\citenamefont {Ekman}\ \emph
  {et~al.}(2021{\natexlab{b}})\citenamefont {Ekman}, \citenamefont {Heinzl},\
  and\ \citenamefont {Ilderton}}]{Ekman:2021vwg}%
  \BibitemOpen
  \bibfield  {author} {\bibinfo {author} {\bibfnamefont {R.}~\bibnamefont
  {Ekman}}, \bibinfo {author} {\bibfnamefont {T.}~\bibnamefont {Heinzl}}, \
  and\ \bibinfo {author} {\bibfnamefont {A.}~\bibnamefont {Ilderton}},\
  }\href@noop {} {\bibfield  {journal} {\bibinfo  {journal} {New J. Phys.}\ }
  (\bibinfo {year} {2021}{\natexlab{b}})},\ \Eprint
  {http://arxiv.org/abs/2102.11843} {arXiv:2102.11843 [hep-ph]} \BibitemShut
  {NoStop}%
\bibitem [{\citenamefont {\'Alvarez}\ and\ \citenamefont
  {Silverstone}(2017)}]{Alvarez:2017sza}%
  \BibitemOpen
  \bibfield  {author} {\bibinfo {author} {\bibfnamefont {G.}~\bibnamefont
  {\'Alvarez}}\ and\ \bibinfo {author} {\bibfnamefont {H.~J.}\ \bibnamefont
  {Silverstone}},\ }\href {\doibase 10.1088/2399-6528/aa8540} {\bibfield
  {journal} {\bibinfo  {journal} {J. Phys. Comm.}\ }\textbf {\bibinfo {volume}
  {1}},\ \bibinfo {pages} {025005} (\bibinfo {year} {2017})},\ \Eprint
  {http://arxiv.org/abs/1706.00329} {arXiv:1706.00329 [math-ph]} \BibitemShut
  {NoStop}%
\bibitem [{\citenamefont {Zhang}(2013)}]{Zhang:2013ria}%
  \BibitemOpen
  \bibfield  {author} {\bibinfo {author} {\bibfnamefont {S.}~\bibnamefont
  {Zhang}},\ }\href {\doibase 10.1093/ptep/ptt099} {\bibfield  {journal}
  {\bibinfo  {journal} {Prog. Theor. Exp. Phys}\ }\textbf {\bibinfo {volume}
  {2013}},\ \bibinfo {pages} {123A01} (\bibinfo {year} {2013})},\ \Eprint
  {http://arxiv.org/abs/1303.7120} {arXiv:1303.7120 [hep-th]} \BibitemShut
  {NoStop}%
\bibitem [{\citenamefont {Borinsky}\ \emph {et~al.}(2021)\citenamefont
  {Borinsky}, \citenamefont {Dunne},\ and\ \citenamefont
  {Meynig}}]{Borinsky:2021hnd}%
  \BibitemOpen
  \bibfield  {author} {\bibinfo {author} {\bibfnamefont {M.}~\bibnamefont
  {Borinsky}}, \bibinfo {author} {\bibfnamefont {G.~V.}\ \bibnamefont {Dunne}},
  \ and\ \bibinfo {author} {\bibfnamefont {M.}~\bibnamefont {Meynig}},\ }\href
  {\doibase 10.3842/SIGMA.2021.087} {\bibfield  {journal} {\bibinfo  {journal}
  {SIGMA}\ }\textbf {\bibinfo {volume} {17}},\ \bibinfo {pages} {087} (\bibinfo
  {year} {2021})},\ \Eprint {http://arxiv.org/abs/2104.00593} {arXiv:2104.00593
  [hep-th]} \BibitemShut {NoStop}%
\bibitem [{\citenamefont {Kleinert}\ and\ \citenamefont
  {Schulte-Frohlinde}(2001)}]{Kleinert:2001ax}%
  \BibitemOpen
  \bibfield  {author} {\bibinfo {author} {\bibfnamefont {H.}~\bibnamefont
  {Kleinert}}\ and\ \bibinfo {author} {\bibfnamefont {V.}~\bibnamefont
  {Schulte-Frohlinde}},\ }\href {\doibase 10.1142/4733} {\emph {\bibinfo
  {title} {{Critical properties of $\phi^4$-theories}}}}\ (\bibinfo
  {publisher} {World Scientific},\ \bibinfo {year} {2001})\BibitemShut
  {NoStop}%
\bibitem [{\citenamefont {Le~Guillou}\ and\ \citenamefont
  {Zinn-Justin}(2012)}]{le2012large}%
  \BibitemOpen
  \bibinfo {editor} {\bibfnamefont {J.-C.}\ \bibnamefont {Le~Guillou}}\ and\
  \bibinfo {editor} {\bibfnamefont {J.}~\bibnamefont {Zinn-Justin}},\ eds.,\
  \href@noop {} {\emph {\bibinfo {title} {Large-order behaviour of perturbation
  theory}}}\ (\bibinfo  {publisher} {Elsevier},\ \bibinfo {year}
  {2012})\BibitemShut {NoStop}%
\bibitem [{\citenamefont {Florio}(2020)}]{Florio:2019hzn}%
  \BibitemOpen
  \bibfield  {author} {\bibinfo {author} {\bibfnamefont {A.}~\bibnamefont
  {Florio}},\ }\href {\doibase 10.1103/PhysRevD.101.013007} {\bibfield
  {journal} {\bibinfo  {journal} {Phys. Rev. D}\ }\textbf {\bibinfo {volume}
  {101}},\ \bibinfo {pages} {013007} (\bibinfo {year} {2020})},\ \Eprint
  {http://arxiv.org/abs/1911.03489} {arXiv:1911.03489 [hep-th]} \BibitemShut
  {NoStop}%
\bibitem [{\citenamefont {Costin}\ and\ \citenamefont
  {Dunne}(2021)}]{Costin:2021bay}%
  \BibitemOpen
  \bibfield  {author} {\bibinfo {author} {\bibfnamefont {O.}~\bibnamefont
  {Costin}}\ and\ \bibinfo {author} {\bibfnamefont {G.~V.}\ \bibnamefont
  {Dunne}},\ }\href {\doibase 10.1140/epjs/s11734-021-00267-x} {\bibfield
  {journal} {\bibinfo  {journal} {Eur. Phys. J. ST}\ }\textbf {\bibinfo
  {volume} {230}},\ \bibinfo {pages} {2679} (\bibinfo {year} {2021})},\ \Eprint
  {http://arxiv.org/abs/2108.01145} {arXiv:2108.01145 [hep-th]} \BibitemShut
  {NoStop}%
\bibitem [{\citenamefont {Mera}\ \emph {et~al.}(2018)\citenamefont {Mera},
  \citenamefont {Pedersen},\ and\ \citenamefont {Nikoli\'c}}]{Mera:2018qte}%
  \BibitemOpen
  \bibfield  {author} {\bibinfo {author} {\bibfnamefont {H.}~\bibnamefont
  {Mera}}, \bibinfo {author} {\bibfnamefont {T.~G.}\ \bibnamefont {Pedersen}},
  \ and\ \bibinfo {author} {\bibfnamefont {B.~K.}\ \bibnamefont {Nikoli\'c}},\
  }\href {\doibase 10.1103/PhysRevD.97.105027} {\bibfield  {journal} {\bibinfo
  {journal} {Phys. Rev. D}\ }\textbf {\bibinfo {volume} {97}},\ \bibinfo
  {pages} {105027} (\bibinfo {year} {2018})},\ \Eprint
  {http://arxiv.org/abs/1802.06034} {arXiv:1802.06034 [hep-th]} \BibitemShut
  {NoStop}%
\bibitem [{\citenamefont {{Stokes}}(1851)}]{Stokes1851}%
  \BibitemOpen
  \bibfield  {author} {\bibinfo {author} {\bibfnamefont {G.~G.}\ \bibnamefont
  {{Stokes}}},\ }\href@noop {} {\bibfield  {journal} {\bibinfo  {journal}
  {Trans. Camb. Phil. Soc.}\ }\textbf {\bibinfo {volume} {9}},\ \bibinfo
  {pages} {166} (\bibinfo {year} {1851})}\BibitemShut {NoStop}%
\bibitem [{\citenamefont {Stokes}(1864)}]{Stokes1864}%
  \BibitemOpen
  \bibfield  {author} {\bibinfo {author} {\bibfnamefont {G.~G.}\ \bibnamefont
  {Stokes}},\ }\href@noop {} {\bibfield  {journal} {\bibinfo  {journal} {Trans.
  Camb. Phil. Soc.}\ }\textbf {\bibinfo {volume} {10}},\ \bibinfo {pages} {105}
  (\bibinfo {year} {1864})}\BibitemShut {NoStop}%
\bibitem [{\citenamefont {Spohn}(2000)}]{Spohn:1999uf}%
  \BibitemOpen
  \bibfield  {author} {\bibinfo {author} {\bibfnamefont {H.}~\bibnamefont
  {Spohn}},\ }\href {\doibase 10.1209/epl/i2000-00268-x} {\bibfield  {journal}
  {\bibinfo  {journal} {Europhys. Lett.}\ }\textbf {\bibinfo {volume} {50}},\
  \bibinfo {pages} {287} (\bibinfo {year} {2000})},\ \Eprint
  {http://arxiv.org/abs/physics/9911027} {arXiv:physics/9911027} \BibitemShut
  {NoStop}%
\bibitem [{\citenamefont {Rohrlich}(1961)}]{Rohrlich1961}%
  \BibitemOpen
  \bibfield  {author} {\bibinfo {author} {\bibfnamefont {F.}~\bibnamefont
  {Rohrlich}},\ }\href {\doibase 10.1016/0003-4916(61)90028-8} {\bibfield
  {journal} {\bibinfo  {journal} {Ann. Phys. (N. Y.)}\ }\textbf {\bibinfo
  {volume} {13}},\ \bibinfo {pages} {93} (\bibinfo {year} {1961})}\BibitemShut
  {NoStop}%
\bibitem [{\citenamefont {Plass}(1961)}]{Plass:1961zz}%
  \BibitemOpen
  \bibfield  {author} {\bibinfo {author} {\bibfnamefont {G.~N.}\ \bibnamefont
  {Plass}},\ }\href {\doibase 10.1103/RevModPhys.33.37} {\bibfield  {journal}
  {\bibinfo  {journal} {Rev. Mod. Phys.}\ }\textbf {\bibinfo {volume} {33}},\
  \bibinfo {pages} {37} (\bibinfo {year} {1961})}\BibitemShut {NoStop}%
\bibitem [{\citenamefont {Alcaine}\ and\ \citenamefont
  {Llanes-Estrada}(2013)}]{PhysRevE.88.033203}%
  \BibitemOpen
  \bibfield  {author} {\bibinfo {author} {\bibfnamefont {G.~G.}\ \bibnamefont
  {Alcaine}}\ and\ \bibinfo {author} {\bibfnamefont {F.~J.}\ \bibnamefont
  {Llanes-Estrada}},\ }\href {\doibase 10.1103/PhysRevE.88.033203} {\bibfield
  {journal} {\bibinfo  {journal} {Phys. Rev. E}\ }\textbf {\bibinfo {volume}
  {88}},\ \bibinfo {pages} {033203} (\bibinfo {year} {2013})}\BibitemShut
  {NoStop}%
\bibitem [{Note1()}]{Note1}%
  \BibitemOpen
  \bibinfo {note} {An expression involving $\protect \qopname \relax o{sinh}$
  appears in Ref.~\cite {Plass:1961zz} but the approach there is not explicitly
  covariant and only seeks to prove that the non-runaway solution is uniform
  motion.}\BibitemShut {Stop}%
\bibitem [{\citenamefont {Ahmed}\ and\ \citenamefont
  {Dunne}(2017)}]{Ahmed:2017lhl}%
  \BibitemOpen
  \bibfield  {author} {\bibinfo {author} {\bibfnamefont {A.}~\bibnamefont
  {Ahmed}}\ and\ \bibinfo {author} {\bibfnamefont {G.~V.}\ \bibnamefont
  {Dunne}},\ }\href {\doibase 10.1007/JHEP11(2017)054} {\bibfield  {journal}
  {\bibinfo  {journal} {JHEP}\ }\textbf {\bibinfo {volume} {11}},\ \bibinfo
  {pages} {054} (\bibinfo {year} {2017})},\ \Eprint
  {http://arxiv.org/abs/1710.01812} {arXiv:1710.01812 [hep-th]} \BibitemShut
  {NoStop}%
\bibitem [{\citenamefont {Podszus}\ and\ \citenamefont
  {Di~Piazza}(2019)}]{Podszus:2018hnz}%
  \BibitemOpen
  \bibfield  {author} {\bibinfo {author} {\bibfnamefont {T.}~\bibnamefont
  {Podszus}}\ and\ \bibinfo {author} {\bibfnamefont {A.}~\bibnamefont
  {Di~Piazza}},\ }\href {\doibase 10.1103/PhysRevD.99.076004} {\bibfield
  {journal} {\bibinfo  {journal} {Phys. Rev. D}\ }\textbf {\bibinfo {volume}
  {99}},\ \bibinfo {pages} {076004} (\bibinfo {year} {2019})},\ \Eprint
  {http://arxiv.org/abs/1812.08673} {arXiv:1812.08673 [hep-ph]} \BibitemShut
  {NoStop}%
\bibitem [{\citenamefont {Ilderton}(2019)}]{Ilderton:2019kqp}%
  \BibitemOpen
  \bibfield  {author} {\bibinfo {author} {\bibfnamefont {A.}~\bibnamefont
  {Ilderton}},\ }\href {\doibase 10.1103/PhysRevD.99.085002} {\bibfield
  {journal} {\bibinfo  {journal} {Phys. Rev. D}\ }\textbf {\bibinfo {volume}
  {99}},\ \bibinfo {pages} {085002} (\bibinfo {year} {2019})},\ \Eprint
  {http://arxiv.org/abs/1901.00317} {arXiv:1901.00317 [hep-ph]} \BibitemShut
  {NoStop}%
\bibitem [{Note2()}]{Note2}%
  \BibitemOpen
  \bibinfo {note} {It is not necessary to normalise to specifically $p_0$; any
  on-shell momentum will do. This choice is the most convenient for the present
  purpose, though. Other choices imply initial conditions other than~\protect
  \cref {eq:instanton-ic-acc,eq:instanton-ic-mom}.}\BibitemShut {Stop}%
\bibitem [{\citenamefont {Heintzmann}\ and\ \citenamefont
  {Grewing}(1972)}]{Heintzmann:1972mn}%
  \BibitemOpen
  \bibfield  {author} {\bibinfo {author} {\bibfnamefont {H.}~\bibnamefont
  {Heintzmann}}\ and\ \bibinfo {author} {\bibfnamefont {M.}~\bibnamefont
  {Grewing}},\ }\href {\doibase 10.1007/BF01386985} {\bibfield  {journal}
  {\bibinfo  {journal} {Z. Phys.}\ }\textbf {\bibinfo {volume} {251}},\
  \bibinfo {pages} {77} (\bibinfo {year} {1972})}\BibitemShut {NoStop}%
\bibitem [{Note3()}]{Note3}%
  \BibitemOpen
  \bibinfo {note} {The pole indicates that there is a minimum lightfront time
  in the past at which the particle was at the speed of light, cf.~Refs.~\cite
  {Tomaras:2000ag,Woodard:2001hi,Ekman:2021vwg}}\BibitemShut {NoStop}%
\bibitem [{\citenamefont {Lipatov}(1977)}]{Lipatov:1977hj}%
  \BibitemOpen
  \bibfield  {author} {\bibinfo {author} {\bibfnamefont {L.~N.}\ \bibnamefont
  {Lipatov}},\ }\href@noop {} {\bibfield  {journal} {\bibinfo  {journal} {JETP
  Lett.}\ }\textbf {\bibinfo {volume} {25}},\ \bibinfo {pages} {104} (\bibinfo
  {year} {1977})}\BibitemShut {NoStop}%
\bibitem [{\citenamefont {Mironov}\ and\ \citenamefont
  {Fedotov}(2021)}]{Mironov:2021bmp}%
  \BibitemOpen
  \bibfield  {author} {\bibinfo {author} {\bibfnamefont {A.~A.}\ \bibnamefont
  {Mironov}}\ and\ \bibinfo {author} {\bibfnamefont {A.~M.}\ \bibnamefont
  {Fedotov}},\ }\href@noop {} {\  (\bibinfo {year} {2021})},\ \Eprint
  {http://arxiv.org/abs/2109.00634} {arXiv:2109.00634 [hep-th]} \BibitemShut
  {NoStop}%
\bibitem [{\citenamefont {Tomaras}\ \emph {et~al.}(2000)\citenamefont
  {Tomaras}, \citenamefont {Tsamis},\ and\ \citenamefont
  {Woodard}}]{Tomaras:2000ag}%
  \BibitemOpen
  \bibfield  {author} {\bibinfo {author} {\bibfnamefont {T.~N.}\ \bibnamefont
  {Tomaras}}, \bibinfo {author} {\bibfnamefont {N.~C.}\ \bibnamefont {Tsamis}},
  \ and\ \bibinfo {author} {\bibfnamefont {R.~P.}\ \bibnamefont {Woodard}},\
  }\href {\doibase 10.1103/PhysRevD.62.125005} {\bibfield  {journal} {\bibinfo
  {journal} {Phys. Rev. D}\ }\textbf {\bibinfo {volume} {62}},\ \bibinfo
  {pages} {125005} (\bibinfo {year} {2000})},\ \Eprint
  {http://arxiv.org/abs/hep-ph/0007166} {arXiv:hep-ph/0007166} \BibitemShut
  {NoStop}%
\bibitem [{\citenamefont {Woodard}(2002)}]{Woodard:2001hi}%
  \BibitemOpen
  \bibfield  {author} {\bibinfo {author} {\bibfnamefont {R.~P.}\ \bibnamefont
  {Woodard}},\ }\href {\doibase 10.1016/S0920-5632(02)01322-1} {\bibfield
  {journal} {\bibinfo  {journal} {Nucl. Phys. B Proc. Suppl.}\ }\textbf
  {\bibinfo {volume} {108}},\ \bibinfo {pages} {165} (\bibinfo {year}
  {2002})},\ \Eprint {http://arxiv.org/abs/hep-th/0111282}
  {arXiv:hep-th/0111282} \BibitemShut {NoStop}%
\end{thebibliography}%
\end{document}